\documentclass[a4paper,11pt]{article}
\pdfoutput=1 % if your are submitting a pdflatex (i.e. if you have
\usepackage{jheppub} % for details on the use of the package, please
\usepackage[T1]{fontenc}
\usepackage[italian,english]{babel}
\usepackage{hyperref}
\usepackage{ifpdf}
\usepackage{subfigure}

\usepackage{amssymb}
\usepackage{amsfonts}
\usepackage{epsf}
\usepackage{rotating}
\usepackage{graphicx}
\usepackage{amsmath}
\usepackage{fancyhdr}
\usepackage{lineno}
\usepackage{babel}
\usepackage{graphics}
\usepackage{pstricks}
\usepackage{color}
\usepackage{multirow}

\newcommand{\nn}{\nonumber}
\newcommand{\lsim}{\mathrel{\mathop{\kern 0pt \rlap
  {\raise.2ex\hbox{$<$}}}
  \lower.9ex\hbox{\kern-.190em $\sim$}}}
\newcommand{\gsim}{\mathrel{\mathop{\kern 0pt \rlap
  {\raise.2ex\hbox{$>$}}}
  \lower.9ex\hbox{\kern-.190em $\sim$}}}

\newcommand{\be}{\begin{equation}}
\newcommand{\ee}{\end{equation}}
\newcommand{\bea}{\begin{eqnarray}}
\newcommand{\eea}{\end{eqnarray}}

\def\etmiss{\not\!\!{E_T}}
\def\ptmiss{\not\!\!{p_T}}
\title{\boldmath Non-standard charged Higgs decay at the LHC in Next-to-Minimal Supersymmetric Standard Model}

\author[a]{Priyotosh Bandyopadhyay}
\author[b]{Katri Huitu}
\author[c]{Saurabh Niyogi}

\affiliation[a]{Dipartimento di Matematica e Fisica "Ennio De Giorgi", \\ Universit\`a del Salento and INFN-Lecce, \\ Via Arnesano, 73100 Lecce, Italy}

\affiliation[b]{Department of Physics, and Helsinki Institute of Physics,\\ P.O.B 64 (Gustaf H{\"a}llstr{\"o}min katu 2), \\FI-00014 University of Helsinki, Finland}

\affiliation[c]{The Institute of Mathematical Sciences,\\ CIT Campus, Chennai, India}

\emailAdd{priyotosh.bandyopadhyay@le.infn.it}
\emailAdd{katri.huitu@helsinki.fi}
\emailAdd{sniyogi@imsc.res.in}

\abstract{We consider next-to-minimal supersymmetric standard model (NMSSM) which has a
gauge singlet superfield. In the scale invariant superpotential we do not have the mass terms and the whole
Lagrangian has an additional $Z_3$ symmetry. This model can have light scalar and/or pseudoscalar allowed by the
recent data from LHC and the old data from LEP. We investigate the situation where a relatively light charged Higgs can decay to
such a singlet-like pseudoscalar and a $W^\pm$ boson giving rise to a final state containing $\tau$ and/or $b$-jets and lepton(s).
Such decays evade the recent bounds on charged Higgs from the LHC, and
according to our PYTHIA-FastJet based simulation can be probed with 10 fb$^{-1}$ 
at the LHC center of mass energy of 13 and 14 TeV.}
\preprint{{~~HIP-2015-40/TH}}

\begin{document}

\maketitle
\flushbottom

\section{Introduction}
The recent discovery of the Higgs boson at the LHC, both by ATLAS \cite{ATLAS} and CMS \cite{CMS, CMS2} is the so-called 
`last key stone' of the Standard Model (SM). The observation of the Higgs boson with mass around $125$ GeV
has reached around or more than $5\sigma$ level for $WW^*$, $ZZ^*$ and
$\gamma\gamma$ modes so far \cite{ ATLAS, CMS, CMS2}. The fermionic modes
are yet to reach the discovery level.
Although this Higgs boson is believed to be responsible 
for the electroweak symmetry breaking (EWSB) mechanism in the SM, precise measurements of its properties
(couplings etc.) are still required to prove this statement.
However, the situation in the Higgs sector
still remains open since various new physics models can explain the presence of the 
newly discovered Higgs boson. 
%
%for presence of  other reprensentations, i.e., singlet and or triplet. 

Any scalar mass not protected by any symmetry leads to the
so-called hierarchy problem \cite{hierarchy} ({\it i.e.} problems in the stability of mass against large
quantum corrections). One of the
most popular solution to this problem is to extend the SM in a minimally
supersymmetric way (called minimal supersymmetric standard model or MSSM). 
%MSSM along with $R$-parity conservation (a $Z_2$ symmetry) gives rise to the lightest  electro-weak super-particle as
%a possible dark matter matter candidate \cite{rparity}. This makes MSSM as a natural extension of SM. 
However, in the CP-conserving sector of the theory
the lightest Higgs mass is bounded from above by the $Z$ mass ($m_{h_1} \lesssim M_Z$).
LEP experiments searched for the supersymmetric Higgs and put a direct lower bound on its mass
to around $93$ GeV \cite{LEPb}.  
Thus to satisfy both the LEP bound and LHC discovery one needs to calculate Higgs
boson mass at one-loop. All the particles that interact with the Higgs boson, 
contribute to its mass via virtual corrections and the dominant ones come from the 
third generation quark and squark sectors due to their large Yukawa coupling with the Higgs. 
In order to achieve $125$ GeV Higgs mass, loop corrections are required to be sizeable in MSSM.
This, in turn, puts strong constraint on the SUSY mass scale.
For the most constrained SUSY scenarios like mSUGRA, the required mass scale is above a few TeV \cite{cMSSMb}. On the other hand, for the 
phenomenological SUSY scenarios, like pMSSM, one either needs large SUSY mass scale or larger mass splitting between the two scalar tops (stop squarks)
\cite{cMSSM}. This, in a sense, brings back the fine-tuning problem. 

Extension of MSSM in a minimal way by adding a singlet scalar superfield is a natural remedy to the problem. 
This scenario is known as next-to-minimal supersymmetric standard model (NMSSM)\cite{Ellwanger:2009dp}
in which a singlet scalar contributes to the Higgs mass at the tree-level as well as at the loop-level. This naturally lifts the Higgs mass to 
the desired range of around $125$ GeV without the requirement of a high mass scale.

NMSSM is originally motivated by solving the  $\mu$-problem in MSSM. The $\mu$-term ($\mu H_u H_d$) 
is the supersymmetric version of the Higgs mass term in SM and provides the mass term for
the higgsino (the fermionic superpartner of Higgs).
It also contributes to the $Z$ boson mass which is certainly at the electroweak scale.
Therefore, one expects it to be of the order of electroweak scale ($\approx 10^2$ GeV to $1$ TeV).
On the other hand, this term is supersymmetry conserving and it could be present at any scale,
assuming practically any value. This leads to the famous $\mu$-problem in MSSM. Introduction of
a singlet scalar superfield which couples to both the Higgs doublets can generate the $\mu$-term dynamically
when the singlet field gets a vacuum expectation value (vev) \cite{nmssm}. Still one can not make the $\mu$-term vanish arbitrarily unless some symmetry prohibits it.
Generally, a discrete symmetry, named $Z_3$ symmetry (which corresponds to multiplication of all components of the chiral superfields
by a phase $e^{2\pi i/3}$) is imposed on the NMSSM superpotential. This discrete symmetry forbids any bilinear term
in chiral superfields, thereby, forcing the $\mu$-term to vanish. In this work we consider a $Z_3$ invariant NMSSM model.
In addition, we can find a light pseudoscalar as a pseudo-Nambu-Goldstone boson (pNGB)
in such a scenario. LEP  searched for the Higgs bosons $h_1$ and $a_1$ via $e^+ e^- \to Z h_1$
and $e^+ e^- \to a_1 h_1$  (in models with multiple Higgs bosons) and their fermionic
decay modes, i.e. $h_1/a_1 \to b \bar{b}, \tau \bar{\tau} $ and $Z \to \ell \bar{\ell}$.
Such light pNGB or otherwise light scalars (both CP even and odd)  when mostly singlet 
couple to the fermions and gauge bosons only via the mixing with the doublet type Higgs bosons,
 and they can evade the LEP bounds \cite{LEPb}.  The singlet type light pseudoscalar is consistent with LHC data. Even if it is not directly produced at the collider, indirect bounds still exist on such hidden (often termed as ``buried'')
state from  Higgs data at the LHC.

Apart from the decay of the discovered Higgs into the hidden scalar/pseudoscalar, the decays of other possible
heavy Higgses can also be interesting. Among them the charged Higgs is very special as it would straightaway
prove the existence of another Higgs doublet or, simpler, an extended Higgs sector. The masses of the other 
Higgs bosons in MSSM ($h_2$, $a$, $h^{\pm}$)
are closely related to each other. For example, the masses of the CP-odd Higgs and charged Higgs bosons are given by
the relation $m_{h^{\pm}}^2 = m^2_{a} + m_{W^{\pm}}^2$ at tree-level. As a result, the decay $h^{\pm} \rightarrow aW^{\pm}$ is not typically possible.
Even with loop corrections, this degeneracy is very unlikely to be broken.
The additional Higgs singlet can play important role here to lift such degeneracy between the charged Higgs and the
CP-odd Higgs. This means that NMSSM has one more CP-odd and one extra CP-even Higgs states compared to MSSM. The two CP-odd states 
mix among each other to give mass eigenstates, and thereby altering the mass relation. Therefore, the lightest CP-odd
state ($a_1$) may become much lighter than the charged Higgs boson thus allowing the decay $h^{\pm} \rightarrow a_1W^{\pm}$.
But kinematics, although an important factor, is not all. The charged Higgs, as we know, is mostly doublet-type. On the other
hand, the lightest pseudoscalar has to have significant singlet component in order to avoid existing collider bounds. Hence, the coupling
$h^{\pm}-a_1-W^{\pm}$ vanishes unless $a_1$ has got some doublet contribution via mixing.

The charged Higgs phenomenology is often considered by comparing its mass to top quark. 
The light charged Higgs scenario corresponds to $m_t > m_{h^{\pm}}$ and the rest is considered the heavy charged Higgs region.
In the first case, the main production process of charged Higgs comes from $pp \rightarrow t \bar{t}$ with top decaying to $b h^+$.
In the same region, the primary decay modes of the charged Higgs are $\tau \nu$ and $c \bar{s}$ (+ their h.c states).
For a charged Higgs heavier than $m_t$, the primary production channel is $pp \rightarrow t h^{\pm}$ and/or $t b h^{\pm}$ 
\cite{Bawa:1989pc,Datta:2001qs}, while the dominant decay mode of $h^{\pm}$ becomes $t \bar{b}$ (+ h.c). However, this region is overflowed by the SM processes
$ttbb$ and $ttZ$ which are difficult to control \cite{Gunion:1993sv}. 
%This is the reason for most of the experimental groups starting from LEP, TeVatron and LHC to take up studies related to light charged Higgs.

In this work, we are interested in studying the phenomenology of a relatively light charged Higgs (of mass just above $m_t$) scenario
 in the framework of NMSSM. 
%NMSSM is a very much viable possibility which is not ruled out yet by Higgs data from LHC. 
We look for a hidden pseudoscalar via the search of the
charged Higgs boson.  In particular, our intention is to establish a probe for 
the charged Higgs boson decaying into a $W^\pm$ and a light singlet-like pesudoscalar which is otherwise
difficult to produce at the LHC. Studies related to a light scalar/pseudoscalar
have been discussed by many authors \cite{NMSSMps}. If these hidden scalars/pseudoscalars have masses $\leq m_{h_{125}}/2$,
they can be explored by the decay channel $h_{125}\to h_1h_1/a_1a_1$. On the other hand, when
the masses are  $> m_{h_{125}}/2$,  the decay channel is no more kinematically allowed. In that case
a light charged Higgs decaying to $a_1/h_1 W^\pm$ may be the next possible option to look for.

For $m_{h^{\pm}} > m_t$, $bg \rightarrow t h^+$ is the best channel
to produce charged Higgses. We focus on a rather non-standard 
decay modes of the charged Higgs: $a_1/h_1 W^\pm$, where the hidden scalars further decay into 
$b$ and/or $\tau$ pairs. We carefully look for different final states based on these non-standard decay 
modes and try to probe such possibilities. Charged Higgs decays to $a_1/h_1 W^\pm$ have previously
been considered in the literature \cite{Rathsman:2012dp,chargedhiggs-pheno}. Particularly, the authors in \cite{Rathsman:2012dp} consider
a similar scenario with light charged Higgs decaying to $a_1 W^\pm$ in NMSSM. They take the usual approach to produce
a $h^{\pm}$ via $t \bar{t}$ production (with one of top quark decaying to $h^+$ and a bottom) and keep their focus on
the region of parameter space where the $a_1$ mass is above the $b \bar{b}$ threshold but still close to
it so that the two b-quarks fragment into a single $b \bar{b}$-jet.

We organise our paper as follows. In Section~\ref{model}
we give a brief introduction to the model. In Section~\ref{scan}
we scan the parameter space considering various theoretical
and experimental bounds and  select some benchmark points in Section~\ref{bnp}.
We perform the collider simulation and present our results in Section~\ref{sim}. Finally we conclude in Section~\ref{conc}.

\section{The Model}\label{model}

 In NMSSM, an $SU(2)_L \otimes U(1)_Y$ singlet complex scalar field $S$ is added to the MSSM Higgs sector.
The extra singlet couples only to the MSSM Higgs doublet. The superpotential contains 
a new singlet interacting with the Higgs doublets along with the well-known Yukawa
interactions of the up and down-type Higgs with the fermions as in the superpotential in MSSM. 
Other dimensionful couplings are forbidden by the imposition of the $Z_3$ symmetry on the superpotential which is 
given as
\begin{equation}
W_{NMSSM}= W_{MSSM}^{\mu=0} + \lambda_S S  H_u \cdot  H_d + \frac{1}{3} \kappa S^3.
\label{spn}
\end{equation}
The dot product denotes the
usual $SU(2)_L$ product: $H_u \cdot  H_d = H_u^{\alpha} \epsilon_{\alpha\beta} H_d^{\beta}$ with
$\epsilon$ being the anti-symmetric matrix with elements off-diag$(1,-1)$.
Note that the bilinear $\mu$-term is generated dynamically once the singlet acquires a vev which breaks the $Z_3$ symmetry.
The effective term $\mu_{\text{eff}} = \lambda \langle S \rangle$ is naturally of the order of
electroweak scale thus solving the supersymmetric $\mu$-problem. 

The tree level scalar potential is given by \cite{Ellwanger:2009dp}
\begin{align}
\begin{split}\label{scpotn}
%    a ={}& b + c + d\\
%         & + e + f + g
       V ={}&  \left( m_{H_{u}}^2 + \lambda^{2} |S|^2 \right) H_{u}^2 + \left( m^2_{H_{d}} + \lambda^{2} |S|^2 \right) H_{d}^2
             + | \lambda H_u \cdot H_d + \kappa S^2 |^2 \\
            & + \frac{g'^2}{8} \left( H_{u}^2 - H_{d}^2 \right)^2 +
           \frac{g^2}{8} \Big[ \left( H_{u}^2 + H_{d}^2 \right)^2 - 4|H_u \cdot H_d|^2 \Big] \\
           & + m_{S}^2 |S|^2 + \Big[ \lambda A_{\lambda} S H_u \cdot H_d + \frac{1}{3} \kappa A_{\kappa} S^3 + h.c. \Big],
\end{split}
\end{align}
where $g^{'}$ and $g$ are the $U(1)_{Y}$ and $SU(2)_{L}$ coupling constants, respectively.
$m_{H_{u}}^2$, $m_{H_{d}}^2$, $m_{S}^2$, $A_{\kappa}$ and $A_{\lambda}$  are the soft breaking 
parameters. We also denote the vev of $H_u$, $H_d$ and $S$ by $v_u$, $v_d$ and $v_s$, respectively,
with the definition $\tan \beta = v_u/v_d$. However, $M_Z$ and $\tan \beta$ 
define $v_u$ and $v_d$. Thus at the tree-level, the Higgs sector in NMSSM
has the following nine parameters:
\begin{equation*}
 \lambda,~\kappa,~\tan \beta, ~ \mu_{\text{eff}},~ A_{\lambda},~ A_{\kappa}, ~ m_{H_{u}}^2,~ m_{H_{d}}^2, ~m_{S}^2.
\end{equation*}
Three minimization conditions corresponding to three scalar superfields in $V$ can fix any three of the parameters. Usually the soft breaking mass
parameters $m_{H_{u}}^2$, $ m_{H_{d}}^2$, $m_{S}^2$ are solved, which leaves six independent parameters.
Out of the ten real degrees of freedom in the fields, three have been used to give masses to the weak gauge bosons
after electroweak symmetry breaking. The other seven become the physical Higgs states with three
CP even ($h'_1$, $h'_2$, $h'_3$ with any one being the $h_{125}$), two CP odd states ($a'_1$, $a'_2$) and
two charged Higgs states ($h^{\pm}$). The neutral CP even Higgs states are given as
\begin{align}\label{cpeven}
 h'_1 &= \sqrt{2} \left( \left( Re~ H_{d}^{0} - v_d \right) \cos \beta -  \left( Re~ H_{u}^{0} - v_u \right) \sin \beta \right) ,\nn \\
 h'_2 &= \sqrt{2} \left( \left( Re~ H_{d}^{0} - v_d \right) \sin \beta +  \left( Re~ H_{u}^{0} - v_u \right) \cos \beta \right) ,\\
 h'_3 &= \sqrt{2} \left( Re~ S - v_s \right) \nn.
\end{align}
% $h_1$ usually behaves like the SM Higgs which has been discovered at the LHC.
The mass matrix may still not be diagonal with these rotations. After diagonalization of the mass matrix, three mass eigenstates, conventionally listed in the order of increasing mass as  $h_1$, $h_2$, $h_3$ for CP-even and $a_1$, $a_2$ for CP-odd,  are mixtures of the gauge (weak) eigenstates.
$h_1$ usually behaves closely like the scalar discovered at the LHC. Similarly, the neutral CP odd Higgses 
and charged Higgs states can be written as,
\begin{align}\label{cpodd}
 a_1 &= \sqrt{2} \left( \left( Im~ H_{d}^{0}  \right) \sin \beta +  \left( Im~ H_{u}^{0} \right) \cos \beta \right)  \nn \\
 a_2 &= \sqrt{2} \left( Im~ S \right),
% h_3 &= \sqrt{2} \left( Re~ S - v_s \right)
\end{align}
and
\begin{align}\label{chhiggs}
 h^{\pm} = H_{d}^{\pm} \sin \beta +  H_{u}^{\pm} \cos \beta .
\end{align}
Charged Higgs states are always purely $SU(2)_L$ doublet states as in MSSM. The compositions of the CP even and odd
Higgs states depend on the parameters. Particularly, $\lambda$ is the main parameter which infuses singlet mixing in the 
CP even and odd Higgses.

%The additional Higgs singlet has interesting consequences on the phenomenology of the Higgs sector.
%In MSSM, the mass of the charged Higgs and CP odd Higgs are related as $m^2_{H^{\pm}} = m^2_A + m^2_{W^{\pm}}$
%at the tree level. As a result, the decay channel $H^{\pm} \rightarrow A W^{\pm}$ is not kinematically 
%allowed. But the extra singlet complex scalar field, introduce one more CP odd Higgs which now mixes 
%with the other CP odd state. Thus one can play around the mass matrix which eventually result in opening the previously forbidden channel.  

\section{Parameter space scan}\label{scan}
 %%%%%%%%%%%%%%%%% Benchmark points%%%%%%%%%%%%%%%%%%%%

 We scan the NMSSM parameter space using the publicly available code NMSSMTools v4.7.0 \cite{nmssmtools}. We focus only on the
 Higgs sector and try to achieve a light pseudoscalar and a relatively light charged Higgs with mass just above the top mass 
 ($\approx 200-250$ GeV) satisfying the LHC Higgs results. We consider the parameter region where BR($h^\pm \to W^\pm a_1$) is significant. 
 Usually the BR($h^\pm \to W^\pm a_1$)  is close to other branching ratios, like  BR($h^{\pm}\to t \bar b$, $\tau \nu$). 
 %has tough competetion from other decays like  BR($h^{\pm}\to t \bar b$, $\tau \nu$). 
 \newline
 \indent
 %Soft mass parameters in the stop sector can, in general, induce large loop corrections to the Higgs masses.
 Here we do not vary the soft mass parameters in the stop sector in order to avoid complicated parameter dependence, but the
 loop corrections in Higgs mass arising from the third generation squarks have been taken into account. We have fixed values for the other slepton and squark masses.
 Soft SUSY breaking gaugino masses are also held to a constant value as they are barely connected to the Higgs sector.
 Though, their values are important while considering decays. The soft-breaking terms are as follows:
 \begin{align*}
&M_1 = 100~\text{GeV,}  ~~M_2 = 200~\text{GeV,} ~~ M_3 = 1.5~\text{TeV,} \\
&M_{L_{1,2}} = 500~\text{GeV} = M_{E_{1,2}}\text{,} ~~ M_{L_{3}} = 500~\text{GeV} = M_{E_{3}}, \\
&M_{Q_{1,2}} = 1.0~\text{TeV} = M_{{U,D_{1,2}}} ,\\
&M_{Q_{3}} = 700~\text{GeV,}  ~ M_{U_{3}} = 900~\text{GeV,} ~ M_{D_{3}} = 800~\text{GeV} ,\\
&A_{b,\tau} = 100~\text{GeV},
 \end{align*}
where $M_{i=1,2,3}$ are the three gaugino masses, $M_{L},M_{E}$ are the doublet and singlet slepton masses, $M_{Q}, M_{U}, M_{D}$ are the doublet and singlet type squark
masses where $i = 1,2,3$ stands for three generations. $A_{b, \tau}$ are trilinear couplings for bottom and tau.

Since our primary goal is to study the Higgs sector, particularly the charged Higgs, we scan over the following parameter region ($\mu , \, M_A,\, A_{\kappa},\, A_t$ are in GeV):
  \begin{eqnarray}
   2.0 ~~~ \leq &~\tan \beta & \leq ~~~ 40.0 \nn \\
   -1000.0 ~~~ \leq & \mu & \leq  ~~~ 1000.0 \nn \\
   0.01 ~~~ \leq & \lambda, \kappa & \leq  ~~~ 1.0 \nn \\
    0 ~~~ \leq & M_A & \leq ~~~ 400.0 \nn \\ 
   -1000.0 ~~~ \leq  & A_{\kappa} & \leq ~~~ 1000.0 \nn\\
   -2000.0 ~~~ \leq &  A_t & \leq ~~~ 2000.0
  \end{eqnarray}
  
   %%%%%%%%%%%%%%%%%%%%%%%%%%%
\begin{figure}[t]
\begin{center}
\includegraphics[width=2.9in,height=2.4in]{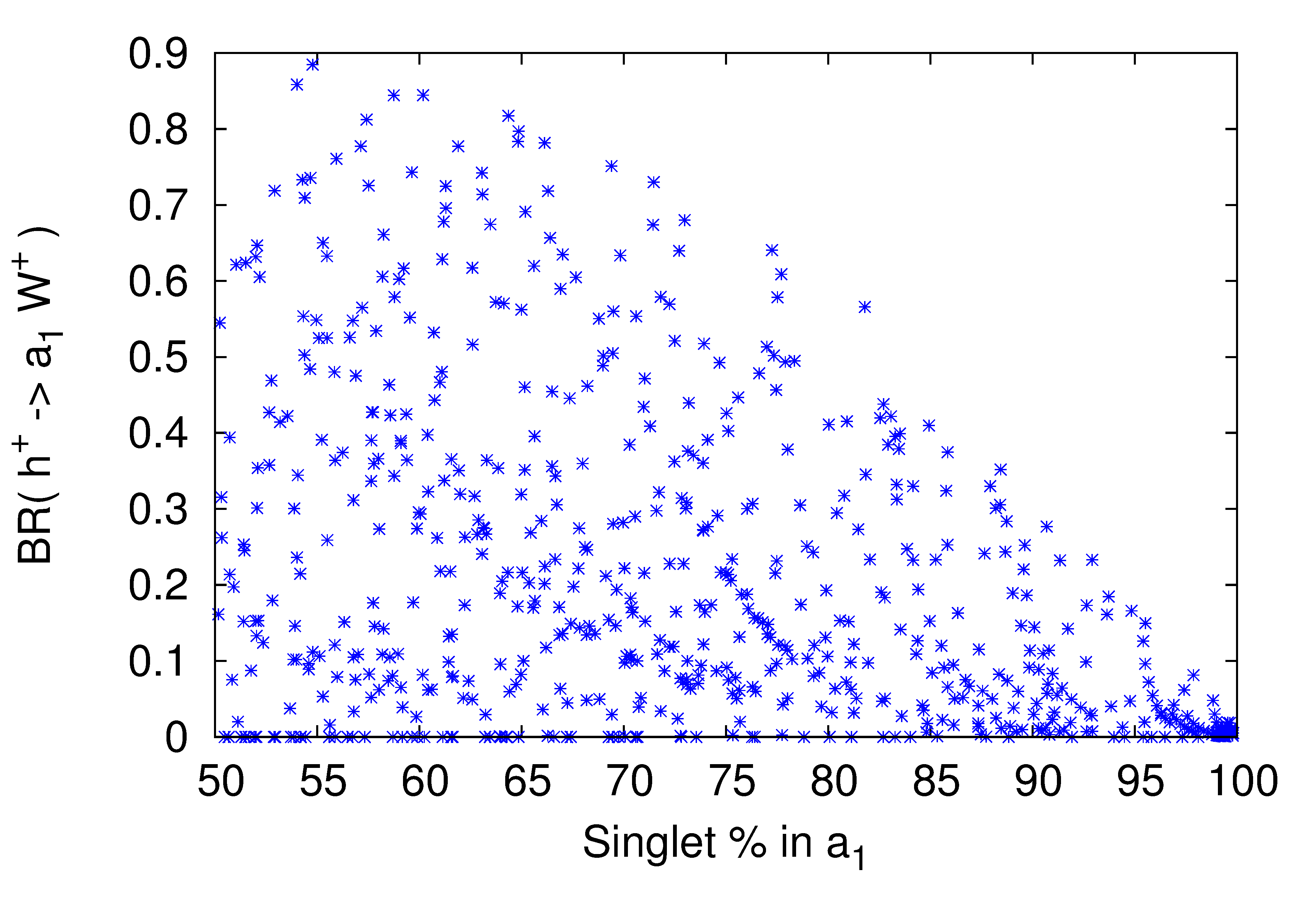}
\includegraphics[width=2.9in,height=2.4in]{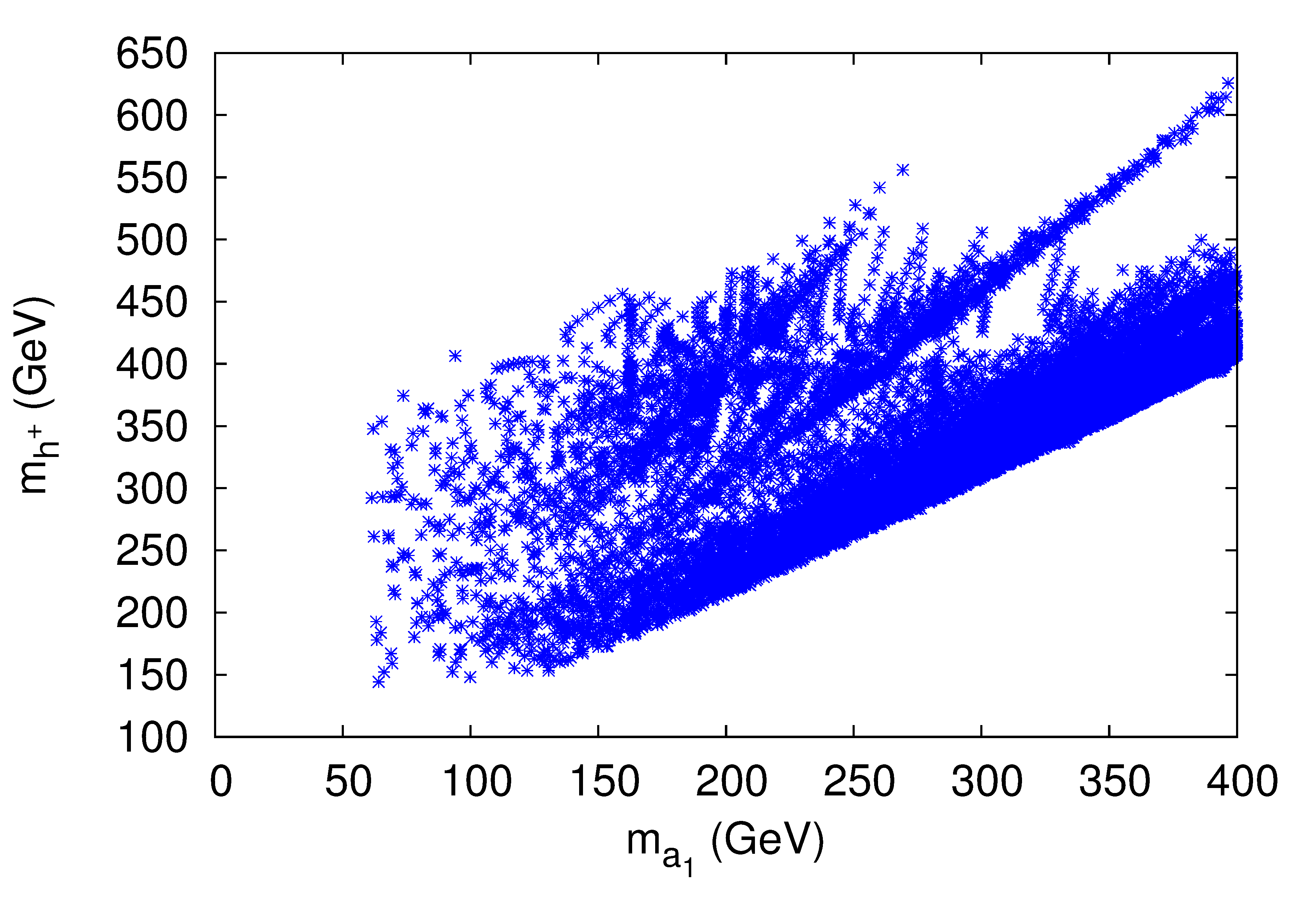}
\caption{Left: Singlet percentage in $a_1$ against BR($h^{\pm} \rightarrow a_1 W^{\pm}$).
         Right: Mass correlation between lightest pseudoscalar and charged Higgs.}\label{fig:fig1}
\end{center}
\end{figure}
%%%%%%%%%%%%%%%%%%%%

%
% %%%%%%%%%%%%%%%%%%%%%%%%
% \begin{figure}[b]
% \centering
% \includegraphics[width=7.5cm,height=6cm]{ch-taunutau.png}
% \includegraphics[width=7.5cm,height=6cm]{ch-tb.png}
% \caption{Charged Higgs branching ratio to $\tau \nu_{\tau}$ and top-bottom against charged Higgs mass. Below the top mass
%          the main decay mode is $\tau \nu_{\tau}$. But as the kinematic region opens for decay to top, it quickly peaks
%          to top-bottom final state.}
% %\label{Serie}
% \end{figure}
%%%%%%%%%%%%%%%%%%%%%%%%%%%
% %%%%%%%%%%%%%%%%%%
% \begin{figure}[t]
% \centering
% \includegraphics[width=10cm,height=6cm]{ch-wa1h1.png}
% \caption{Charged Higgs branching fraction to two competing modes}
% %\label{Serie}
% \end{figure}
%%%%%%%%%%%%%%%%%%

\noindent
 The SUSY scale is fixed at $M_{SUSY}$ = 1 TeV. We demand that the lightest Higgs, $h_1$,
 lies within 122-128 GeV. We also require the lighter chargino mass to be greater than 105 GeV to avoid the LEP bound from
 chargino pair production \cite{chargino}.

 Left panel of Fig.~\ref{fig:fig1} shows the variation of  $h^{\pm} \rightarrow a_1 W^{\pm}$ branching ratio 
 with percentage of singlet component in the lightest pseudoscalar $a_1$. As mentioned in the introduction, 
 the lightest pseudoscalar must be singlet-like to evade the LEP bound. On the other hand, $h^{\pm}$ is only
 doublet-type. Therefore to have the desired decay mode, one must have enough doublet component in $a_1$ via
 mixing. When $a_1$ becomes a pure singlet, the branching ratio goes to 0. The right plot
 is the correlation between masses of $a_1$ and $h^{\pm}$. We notice the diagonal behaviour
 which is clearly the MSSM limit. These two figures together give us an idea about the dynamics ({\it i.e.} coupling)
 and the kinematics between charged Higgs and the lightest pseudoscalar to maximize the corresponding decay channel.

\begin{figure}[t]
\begin{tabular}{cc}
\includegraphics[width=.5\linewidth]{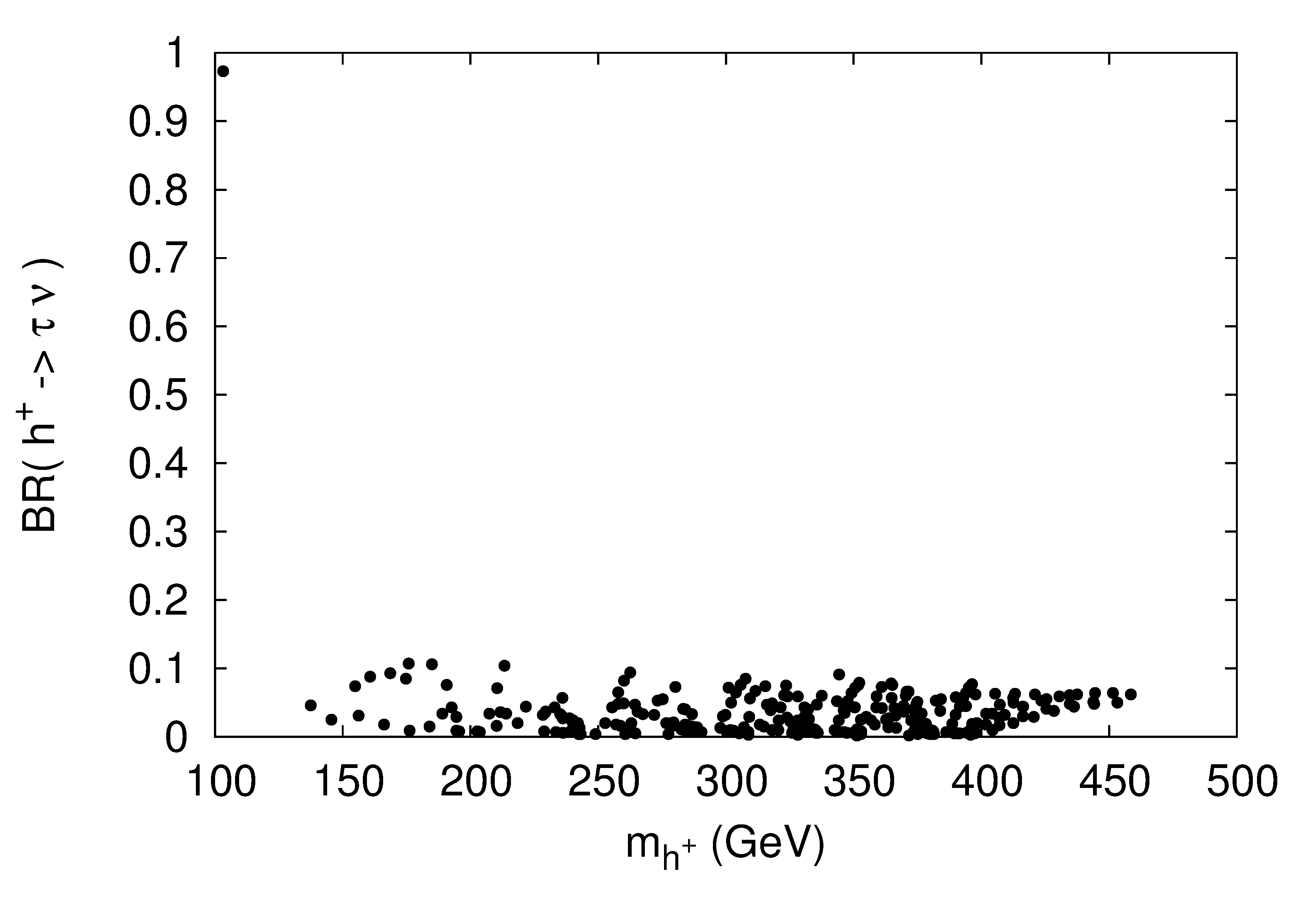}&
\includegraphics[width=.5\linewidth]{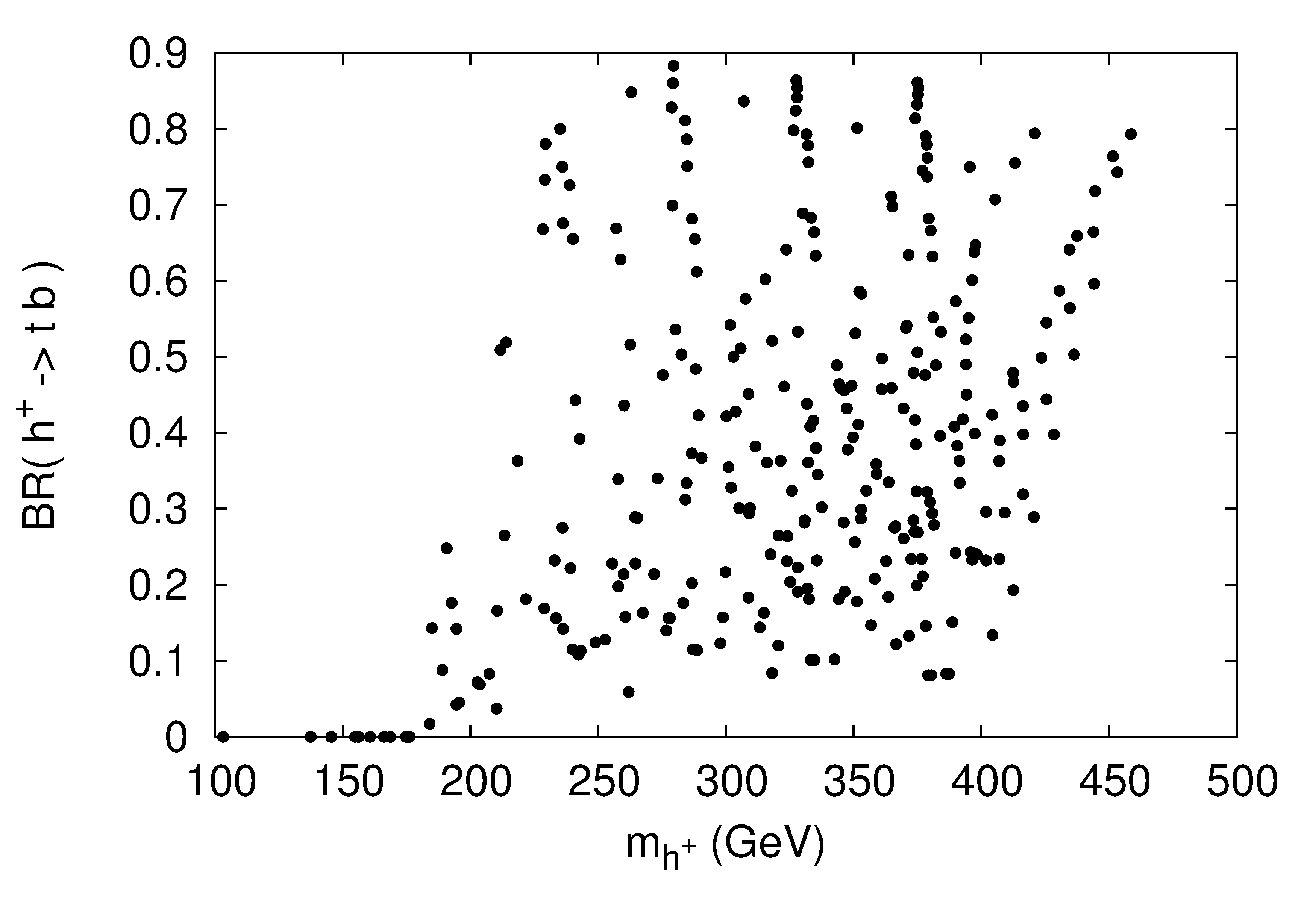} \\
\includegraphics[width=.5\linewidth]{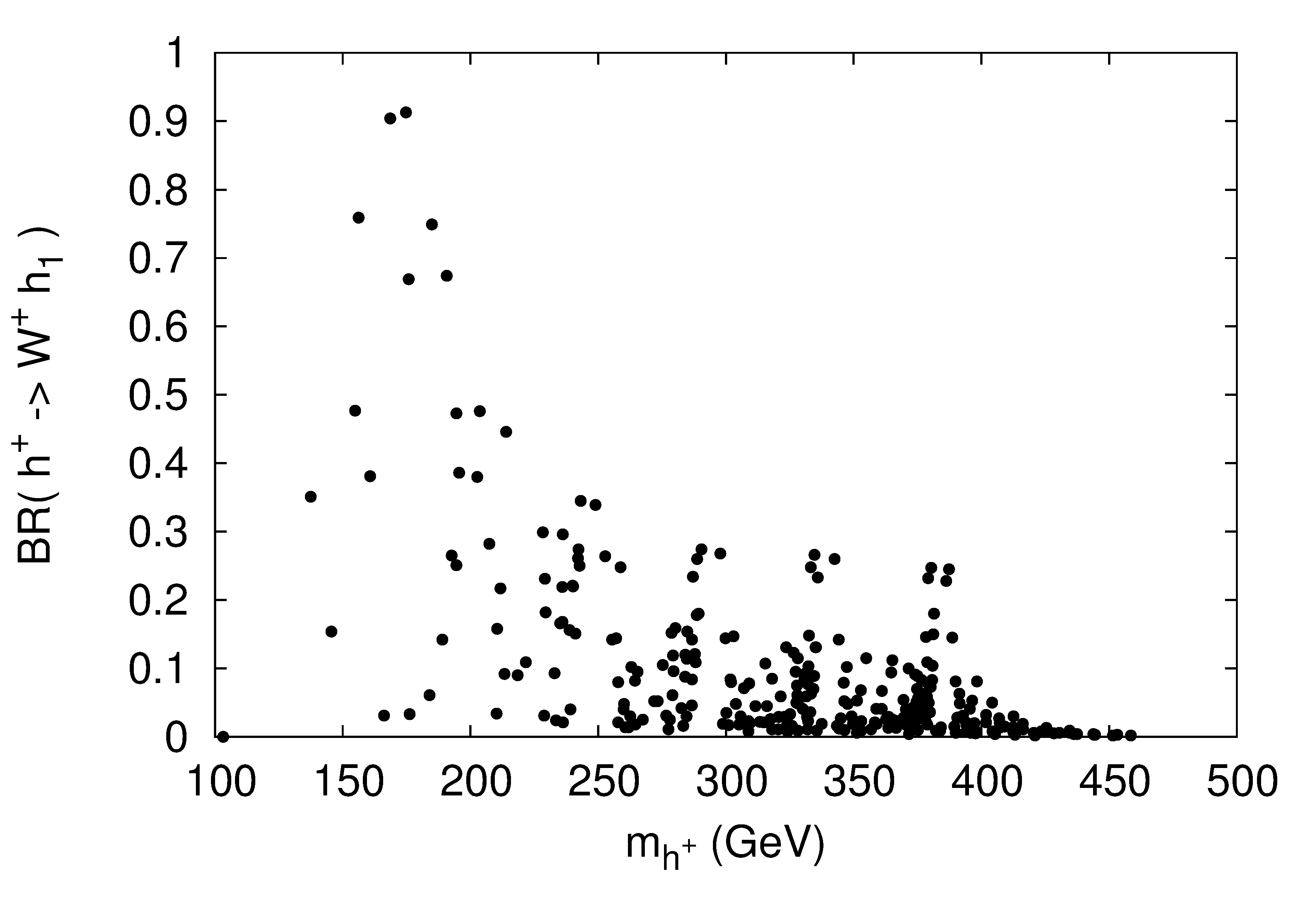}&
\includegraphics[width=.5\linewidth]{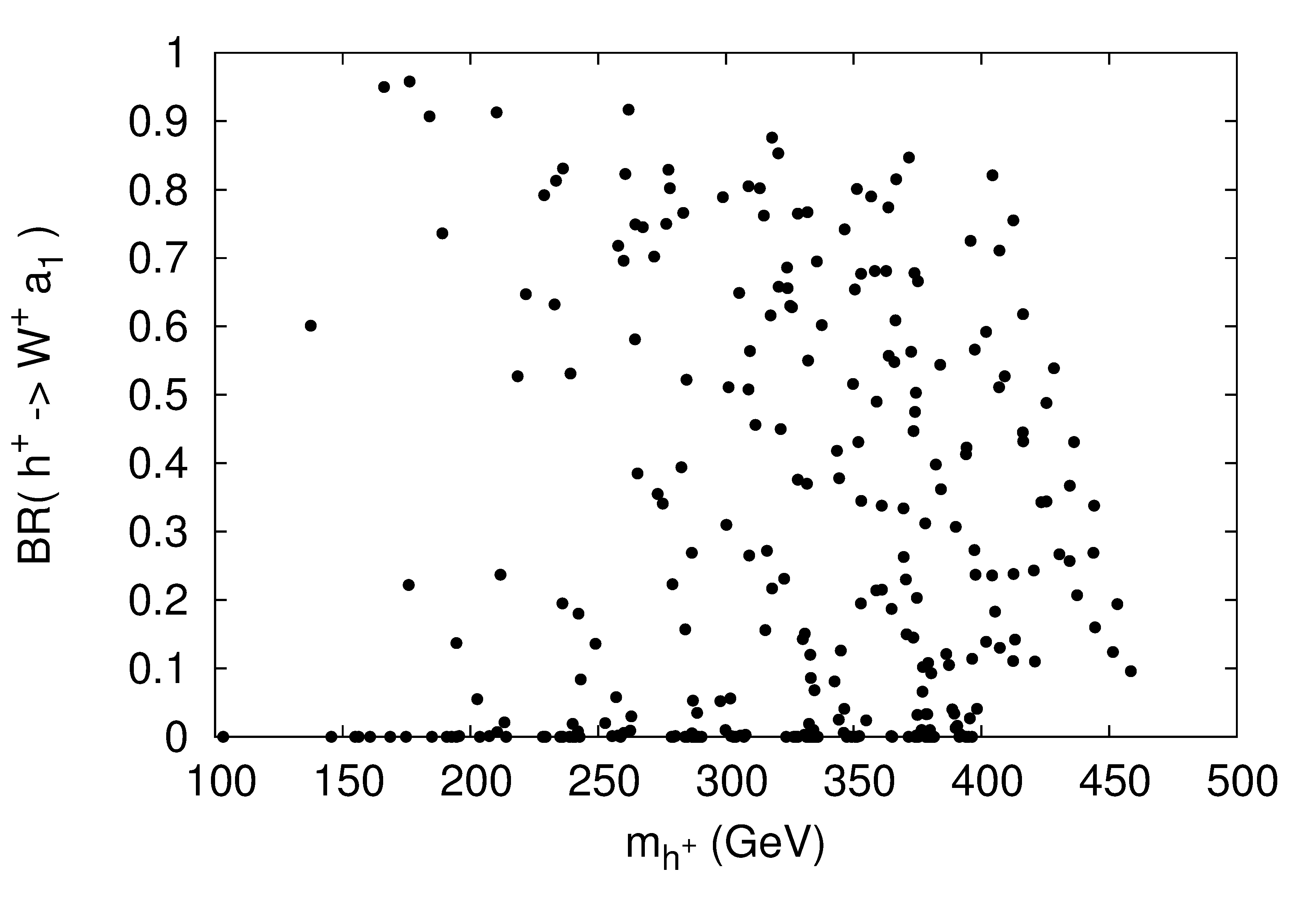}
\end{tabular}
\caption{Top row: Charged Higgs branching ratio to $\tau \nu$ and $t b$ against $m_{h^{+}}$.
         Bottom row : Same with $W^{+} h_{1}$ and $W^{+} a_{1}$.}\label{chbr}
\end{figure}

 In Fig.~\ref{chbr} we show charged Higgs branching fractions in various important decay channels. As we can see, 
$\tau \nu$ channel is dominant for the lightest charged Higgs masses. Once $m_{h^+}$ hits the top threshold, the $t \bar{b}$ channel becomes
the dominant decay mode and $\tau \nu$ mode remains within 10\% for the rest of the region. $W^{\pm} h_1$ also 
remains substantial in the desired charged Higgs mass range of $\approx 200-250$ GeV. Branching ratio for charged Higgs decaying to 
$W^{\pm} a_{1}$ can be quite large ($\sim$ 60-70\%) for light $h^{\pm}$. As mass increases, the $t \bar{b}$ decay mode
becomes significant. Interesting to note that in the first three plots starting from top-left,
the masses of the decay products are known. Still, we see the branching ratios to be varying even for fixed mass of 
the parent particle. This is simply because most of the couplings depend on $\tan \beta$ (or, $ \cot \beta$) and 
the mixing angle in the Higgs sector. Variation in relevant parameters keep the couplings changing. On the other hand,
$a_1$ mass and its coupling in the bottom-right plot is varying. 
The scatter plots in Fig.~\ref{chbr} show how the charged Higgs branching ratios to various channels vary over its mass,
although the mass of the charged Higgs is fixed.

%The scatter plots are just to show
%how the charged Higgs branching ratios to various channels vary over its mass. With these plots one can see 
%different available branching fractions for a particular decay mode even for a fixed mass of the charged Higgs. 

%%%%%%%%%%%%%%%%%%%

\section{Benchmark points}\label{bnp}
In this section we carefully select three points which satisfy the recent bounds from
LHC \cite{ATLAS,CMS,CMS2} and LEP \cite{LEPb} to carry out the collider study.
The parameters and the resulting mass spectrum for the chosen benchmark points are given in Table~\ref{param}
and \ref{bps}, respectively. BP1 and BP2 have larger values for $\lambda, \kappa \gsim 0.75$ than BP3, and the theory hits Landau  pole before the GUT scale \cite{Ellwanger:2009dp}. All the three benchmark points have a
light, mostly singlet pseudoscalar with mass around 65-75 GeV, and is yet to be found at LHC and is allowed by
the LEP data. We avoid $h_1 \rightarrow a_1 a_1$ mode by choosing $m_{a_1} > \frac{m_{h_1}}{2}$,  although
BR$(h_{125} \to a_1 a_1)\lesssim 20\%$ is still allowed with the current uncertainty \cite{hiddenHE}.
The charged Higgs spectrum is relatively light 
with the lightest one for BP3 around 182 GeV and heaviest one for BP1, around 250 GeV. For all the
three benchmark points the lightest CP-even Higgs eigenstate is the discovered scalar at the LHC which satisfies the Higgs data within $2\sigma$ of the signal strengths $\mu=\frac{\sigma(pp\to h)\times\mathcal{B}(h\to XX)}{\sigma(pp\to h)_{\rm{SM}}\times\mathcal{B}(h\to XX)_{\rm{SM}}}$  in $WW^*$, $ZZ^*$ and $\gamma\gamma$ modes from ATLAS \cite{ATLAS} and CMS \cite{CMS, CMS2}. 
We have also taken into account the recent bounds on the third generation squarks from the LHC \cite{thridgensusy} and demanded the 
lighter chargino to be heavier than $105$ GeV.
%
%%%%%  tablle -1 
\begin{table}
\begin{center}
\renewcommand{\arraystretch}{1.4}
\begin{tabular}{||c|c|c|c||}
\hline\hline
Parameters & BP1 & BP2 & BP3\\
 & & &\\ \hline\hline
$\tan \beta$ & 3.0 & 2.0 & 40.0 \\
\hline
$ \lambda $ & 0.75 & 0.88  & 0.26 \\
\hline
$ \kappa $ & 0.90 & 0.88  & 0.51 \\
\hline
$ A_{\kappa}$& -60.0 & 100.0 &  -100.0 \\
\hline
$ M_{A}$ & 270.9 & 245.7 & 280.0 \\
\hline
$ \mu $ & -102.0 & -200.0 & 190.0 \\
\hline
$ A_{t}$ & 75.0 & 100.0 & 1500.0   \\
\hline
\hline
\end{tabular}
\caption{Parameter sets chosen as the benchmark points for the collider analysis.
The mass spectra are given below in Table~\ref{bps}. }\label{param}
\end{center}
\end{table}

%
%%%%%%%%%%%%%%%%%%%%%%%%%%%   table -2 
\begin{table}
\begin{center}
\renewcommand{\arraystretch}{1.4}
\begin{tabular}{||c|c|c|c||}
\hline\hline
Benchmark & BP1 & BP2 & BP3\\
Points & & &\\ \hline\hline
$m_{h_1}$ & 123.9 & 123.88 & 123.67 \\
\hline
$m_{h_2}$ & 185.9 & 218.9  & 169.67 \\
\hline
$m_{h_3}$ & 321.5 & 374.13  & 717.27 \\
\hline
\hline
$m_{a_1}$& 73.8 & 65.99 &  71.38 \\
\hline
$m_{a_2}$ & 277.5 & 375.05 & 362.48 \\
\hline
\hline
$m_{h^\pm}$ & 250.3 & 212.05 & 182.4 \\
\hline
%$m_{h^\pm_2}$ & &  &  \\
\hline
%\hline
$m_{\tilde{t}_1}$ & 747.18 & 745.63 & 644.14   \\
\hline
$m_{\tilde{t}_2}$ & 944.97 & 945.01 & 980.54  \\
\hline
$m_{\tilde{b}_1}$& 734.93 & 733.89 &  719.32 \\
\hline
$m_{\tilde{b}_2}$& 835.99 & 835.77  & 834.89   \\
\hline
\hline
\end{tabular}
\caption{Particle spectra for our benchmark points.}\label{bps}
%consistent with the recent LHC data. }\label{bps}
\end{center}
\end{table}
%%%%%%%%%%%%  

%%%%%%%%%%%%%%%%% h_125 decay branching fraction (with tree level mass)%%%%%%%%%%%%%%%%%%%%
\begin{table}
\begin{center}
\renewcommand{\arraystretch}{1.4}
\begin{tabular}{||c|c|c|c|c|c|c||}
\hline\hline
Benchmark&\multicolumn{4}{|c||}{Branching fractions}\\
\cline{2-5}
Points &\; $W^+ W^-$ \;  &\; $Z Z$ \;& \;$b\bar{b}$ \;&\;$\tau \bar{\tau}$ \\
\hline\hline
BP1 & 23.4\% & 2.8\% & 56.2\% & 5.83\% \\
\hline
BP2 &  39.8\% & 4.89\% & 26.99\% & 2.66\% \\
\hline
BP3 & 12.39\%  & 1.52\% &  72.12\% & 8.25\% \\
\hline
\hline
\end{tabular}
\caption{Some major decay branching fractions of $h_{1}$ for the benchmark points.}\label{hdcy2}
\end{center}
\end{table}
%%%%%%%%%%%%

%%%%%%%%%%%%%%%%% a_1 decay branching fraction (with tree level mass)%%%%%%%%%%%%%%%%%%%%
\begin{table}
\begin{center}
\renewcommand{\arraystretch}{1.4}
\begin{tabular}{||c|c|c|c|c|c|c|c|c||}
\hline\hline
Benchmark&\multicolumn{2}{|c||}{Branching fractions}\\
\cline{2-3}
Points& \;$b\bar{b}$ \;&\;$\tau \bar{\tau}$ \\
\hline\hline
BP1 & 91.1\% & 8.58\% \\
\hline
BP2& 91.0\% & 8.31\%\\
\hline
BP3& 87.95\%  & 11.69\% \\
\hline\hline
\end{tabular}
\caption{Decay branching fractions of $a_1$ for the benchmark points.} \label{a1dcy2}
%The kinematically forbidden decays are marked with dashes.}\label{a1dcy2}
\end{center}
\end{table}
%%%%%%%%%%%%
Table~\ref{hdcy2} presents some decay branching fractions for $h_1$, which is the discovered scalar around
$\sim 125$ GeV. The dominant decay branching fractions are within $1\sigma$
uncertainties of both ATLAS results \cite{ATLAS} and CMS \cite{CMS, CMS2}. Table~\ref{a1dcy2} presents decay
branching fractions of the light pseudoscalar which dominantly decay to $b\bar{b}$ and $\tau\bar{\tau}$.
From Table~\ref{chdcy2} we see that for the chosen benchmark points the light charged Higgs can decay to $a_1 W^\pm$ along
with the other channels ($\tau \nu$ and $tb$). BR($h^\pm\to a_1 W^\pm$) can be as large as $\sim 66\%$ (for BP3)
which shows that such a non-standard decay mode is very much possible. In the case of BP1 and BP2, $h^\pm\to h_{1} W^\pm$ mode
is open but the branching fraction is rather small ($\lsim 1\%$). In the case 
of BP1, the charged Higgs decaying to lighter chargino is open via $h^\pm \to \tilde{\chi}^\pm_1 \tilde{\chi}^0_{1,2,3}$
with a branching fraction of $\sim 21\%$. The charged Higgs boson decaying into supersymmetric modes could be a good
probe for lighter gauginos and higgsinos at the LHC.

%%%%%%%%%%%%%%%%% h^+_1 decay branching fraction (with tree level mass)%%%%%%%%%%%%%%%%%%%%
\begin{table}
\begin{center}
\renewcommand{\arraystretch}{1.4}
\begin{tabular}{||c|c|c|c|c|c|c|c|c|c||}
\hline\hline
Benchmark&\multicolumn{4}{|c|}{Branching fractions}\\
\cline{2-5}
Points& $h_1W^\pm$ \;& \;$a_1 W^\pm$ \;&\;$\tau \bar{\nu}$& \;$t \bar{b}$  \\
\hline\hline
BP1 & 0.28\% & 18.9\% & 0.59\% & 59.00\%\\
\hline
BP2 & 0.37\%& 65.6\% & 0.17\% & 33.83\%\\
\hline
BP3 & - & 27.43\% & 60.71\% & 10.62\% \\
\hline\hline
\end{tabular}
\caption{Some major Decay branching fractions of $h^\pm$ for the benchmark points.} \label{chdcy2}
%The kinematically forbidden decays are marked with dashes.}
\end{center}
\end{table}
%%%%%%%%%%%% 

\subsection{Production processes}

%%%%%%%%%%%%%%%%%%%%%%%%%%%
 \begin{figure}[t]
\begin{center}
\includegraphics[width=.81\linewidth]{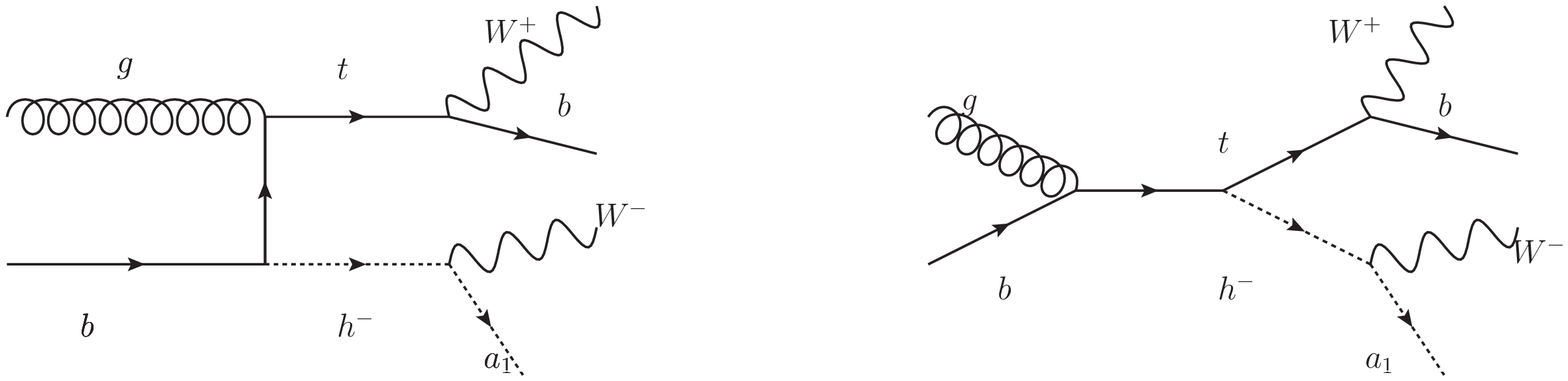}
\caption{Feynman diagrams of $bg\to t h^{\pm}$ production at the LHC.}\label{fig:fy}
\end{center}
\end{figure}
%%%%%%%%%%%%%%%%%%%%

\begin{figure}[t]
\begin{center}
\includegraphics[width=3.2in,height=4.0in,angle=-90]{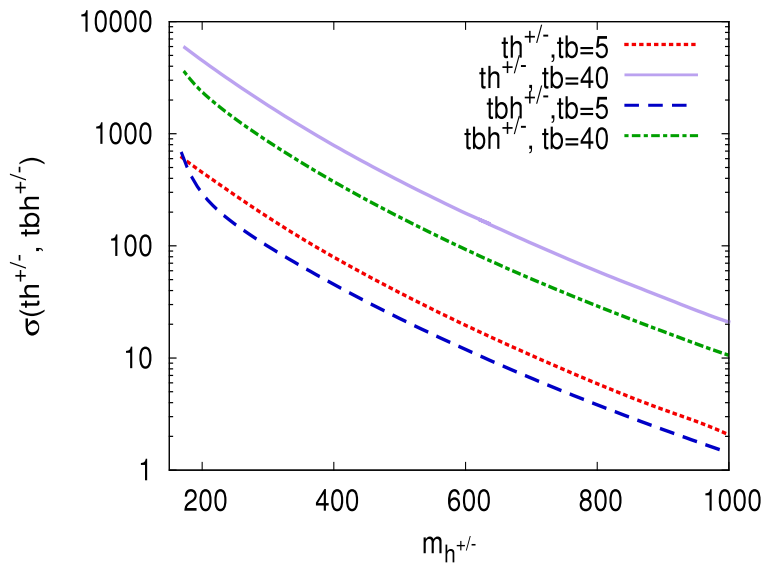}
\caption{Cross section in fb for $pp \to t h^{\pm}$ and $pp\to tb h^\pm$ vs mass of the charged Higgs boson. The blue, green are
for $tbh^\pm$ and red, violet are for $th^\pm$ production processes at ECM=14 TeV for $\tan{\beta}=5, 40$ respectively (see text). The renormalization/factorization
scale is $Q = \sqrt{\hat{s}}$ with CTEQ6L \cite{CTEQ} as PDF. K-factor is taken to be 1.55 \cite{Kidonakis:2008vd}.}\label{cross}
\end{center}
\end{figure}

For our case with $m_{h^{\pm}} > m_t$, the dominant production modes of the charged Higgs
is $bg$ fusion as shown in Fig.~\ref{fig:fy}. In this case 
we produce a single charged Higgs boson in association with a top quark.
% and a top and a bottom quarks.
The other production modes (for {\it e.g.} pair production or $W^{\pm}$/$Z$ associated production) contribute less.
The charged Higgs in NMSSM is exactly same as in MSSM or 2HDM-II. Its coupling to top and bottom quarks has two parts:
one is proportional to $m_t \cot{\beta}$ and the other part is proportional to $m_b \tan{\beta}$. 
This feature makes the top (or bottom) mediated production modes highly  $\tan \beta$ 
dependent as can be seen from figure~\ref{cross}. For BP1 such cross sections are relatively suppressed
 compared to BP2 (relatively lower $\tan \beta$)  and BP3 (high $\tan \beta$) points respectively \cite{Datta:2001qs}.

 The cross sections have been calculated with the renormalization/factorization scale $Q = \sqrt{\hat{s}}$ and with CTEQ6L \cite{CTEQ} as PDF.
The charged Higgs can then decay to a light pseudoscalar and a $W^\pm$ and the top quark decays to
$bW^\pm$ as shown in  Fig.~\ref{fig:fy} thus producing in this case $ 1b + 2W^\pm+ a_1$.
The light  pseudoscalar will further decay into $b$ or $\tau$ pairs.  This will lead to two different 
final states at the parton level $1b + 2 \tau + 2\ell +\etmiss$ and $3b + 2\ell +\etmiss$, if 
both the $W^\pm$ bosons decay leptonically. Table~\ref{prodc} shows the production cross section for the chosen benchmark points. BP3 has the largest cross section
due to enhancement of the Yukawa coupling at high $\tan \beta$. Figure~\ref{cross} shows the variation of 
$pp\to th^\pm$ and $pp\to tb h^\pm$ production cross sections at the LHC with the charged Higgs mass for a given $\tan{\beta}$. The blue dashed
and green dot-dashed lines are for $\sigma(tbh^\pm)$ at 14 TeV for $\tan \beta$ = 5 and 40, respectively. Similarly, 
the red dotted curve and the violet contour are for $\sigma(th^\pm)$ at 14 TeV for $\tan \beta$ = 5 and 40, respectively.

%%%%%%%%%%%
\begin{table}[b]
\begin{center}
\renewcommand{\arraystretch}{1.4}
\begin{tabular}{||c|c|c||}
\hline\hline
Benchmark&\multicolumn{2}{|c|}{Production cross sections (fb)}\\
\cline{2-3}
Points&\;$h^\pm t$&\;$h^\pm t \bar b$ \\
\hline\hline
BP1 & 635.00 (497.26)& 376.73 (303.89)\\
\hline
BP2 & 1433.04 (1169.35)&1206.83 (979.60)\\
\hline
BP3 & 5577.89 (4572.50) & 4482.60 (2421.317)\\
\hline\hline
\end{tabular}
\caption{Charged Higgs production cross sections in association with top quark and top-bottom quarks 
for the benchmark points for 14 (13) TeV. The renormalization/factorization
scale is $Q = \sqrt{\hat{s}}$ with CTEQ6L \cite{CTEQ} as PDF. K-factor is taken to be 1.55 \cite{Kidonakis:2008vd}.}\label{prodc}
\end{center}
\end{table}
%%%%%%%%%%%%
\section{Signature and collider simulation}\label{sim}

%%%%%%%%%%%%%%%%%%%%%%%%%%%
%%%%%%%%%%%%%%%%%%%%

%The discovered Higgs boson around 125 GeV  decays to two light pseudoscalars $<100$ GeV,  is not kinematically allowed
%for the benchmark points. Never the less the charged Higgs bosons for the benchmark points are heavy enough to open the decay 
%mode $h^\pm \to a_1 W^\pm$. The light pseudoscalar produced this way will further decay into tau or $b$ pairs. Possibility
%of such hidden Higgs thus can be found out in the final states with $b$s and $\tau$s.  

We implement the model in SARAH \cite{sarah} and generate the model files for CalcHEP \cite{calchep} which we use to generate the decay
SLHA file. The generated events are then simulated with {\tt PYTHIA} \cite{pythia} via the SLHA interface \cite{slha} for the decay branching and mass spectrum. 
%%%%%%%%%%%%%%%%%%%%%% NJets and jet pt distributions %%%%%%%%%%%%%%%
\begin{figure}
\begin{center}
\includegraphics[width=.33\linewidth, angle=-90]{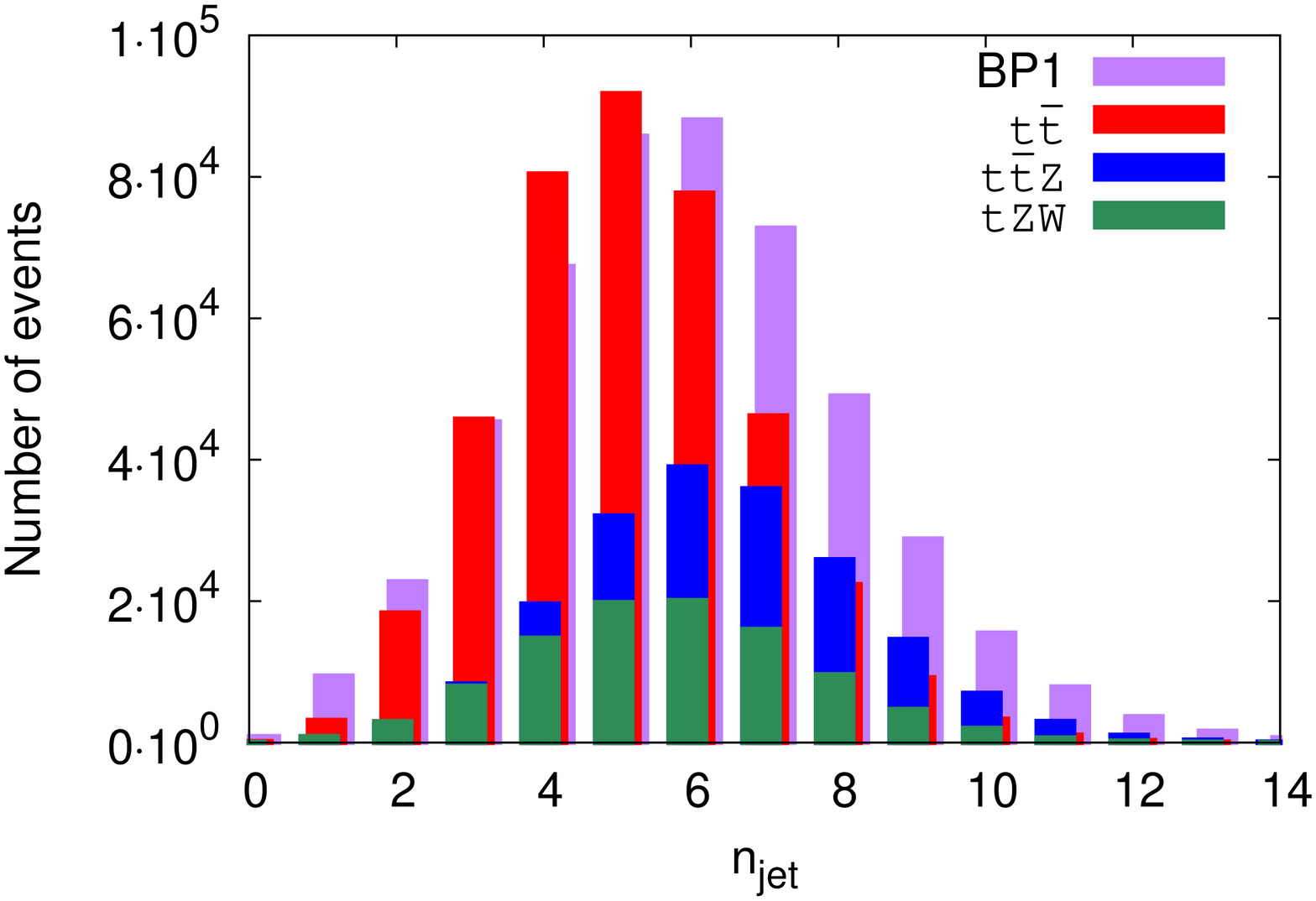}
\includegraphics[width=.33\linewidth, angle=-90]{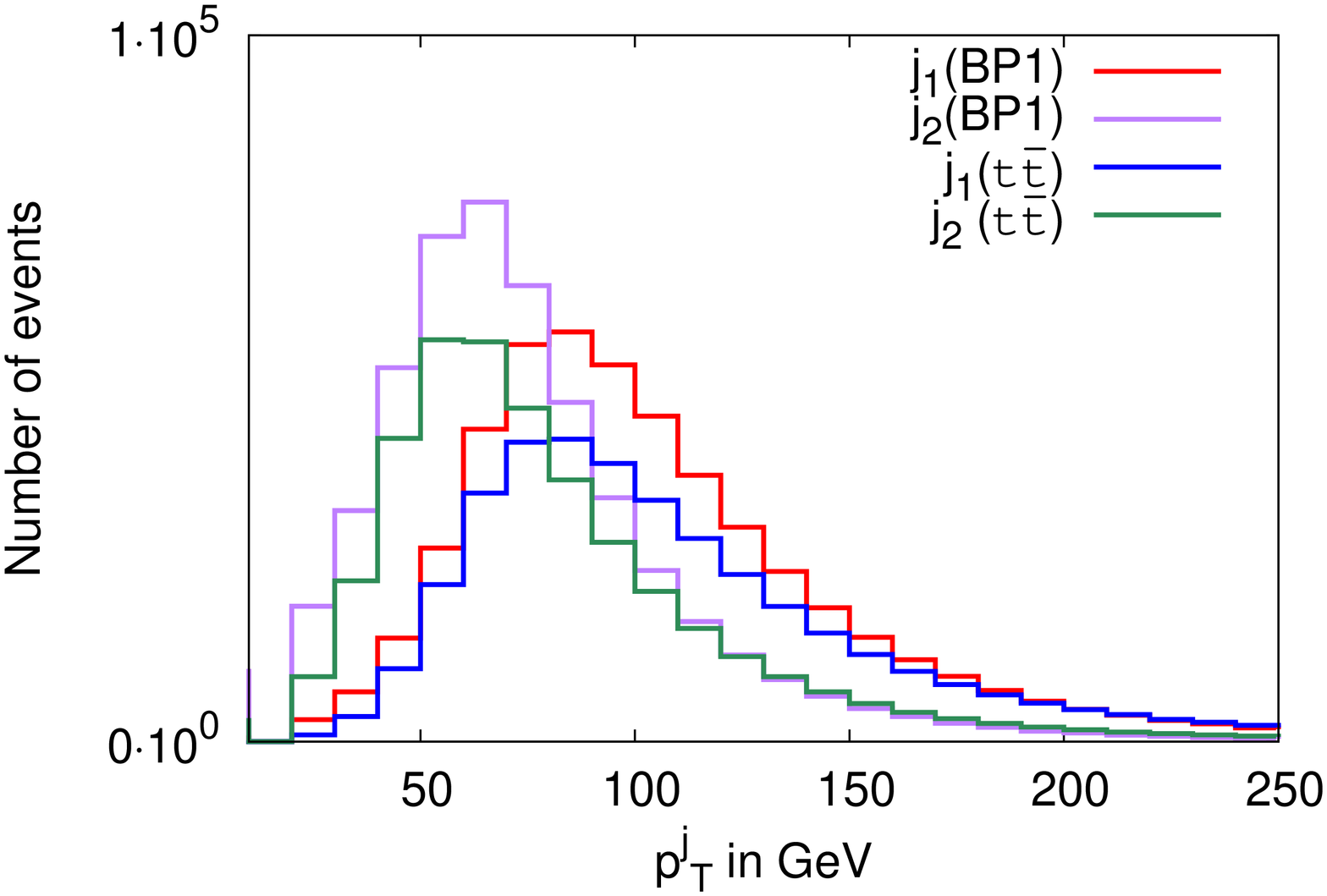}
\caption{Jet multiplicity ($n_j$) distributions (left) for the signal (BP1) and dominant SM backgrounds
$t\bar{t}$, $t\bar{t} Z$ and $t Z W$. Jet $p_T$ distributions (right) of the two hardest jets (descending order in $p_T$)
for signal (BP1) and background $t\bar{t}$.}\label{njjpt}
\end{center}
\end{figure}
%%%%%%%%%%%%%%%%%%%%%%%%%%%%%%%%%%%%%%%%%%%%%%%%%%%%%%%%%%%%%%%%%%%

For hadronic level simulation we have used {\tt Fastjet-3.0.3} \cite{fastjet} package with the {\tt CAMBRIDGE-AACHEN} algorithm \cite{ca-algo}. We have
set a jet size $R=0.5$ for jet formation. We have used the following isolation and selection criteria for leptons and jets:
\begin{itemize}
  \item the calorimeter coverage is $\rm |\eta| < 4.5$
  \item $ p_{T,min}^{jet} = 20$ GeV and jets are ordered in $p_{T}$
  \item leptons ($\rm \ell=e,~\mu$) are selected with
        $p_T \ge 20$ GeV and $\rm |\eta| \le 2.5$
  \item no jet should match with a hard lepton in the event
  \item $\Delta R_{lj}\geq 0.4$ and $\Delta R_{ll}\geq 0.2$
  \item Since efficient identification of the leptons is crucial for our study, we additionally require  
hadronic activity within a cone of $\Delta R = 0.3$ between two isolated leptons to be $\leq 0.15 p^{\ell}_T$ GeV in the specified cone.

\end{itemize}

%The charged Higgs boson will mostly be produced by $b g$ fusion and $pp\to tb h^\pm$ for the benchmark points (see Table~\ref{prodc}). 
%For BP3 $m_{h^\pm} < m_t $, thus the second contributions comes from the on-shell top decay.  The other production modes contribute relatively less. 

%We consider $b g \to t h^\pm$ as the main production process. The charged Higgs boson thus produced decay via $a_1 W^\pm$,
%which is non-standard with respect to the recent charged Higgs searches at the LHC \cite{ChCMS, ChATLAS}.  
We consider $t h^{\pm}$ and $t b h^{\pm}$ (+ h.c.) as the main production channels with charged Higgs decaying to $a_1 W^\pm$.
As discussed earlier, with the subsequent decays these lead to final states with $3b \, + 2W^\pm$ or $1b+2\tau+2W^\pm$.
%The light pseudoscalar further decays into a $b$ or a $\tau$ pair (see Table~\ref{a1dcy2}).
%On the other end, top quark decays into $b W^\pm$. This can lead to two different final states with $3b \, + 2W^\pm$ or $1b+2\tau+2W^\pm$.
%Depending on the $W^\pm$'s decay to leptonic or hadronic mode, the final states will be different. 

Such  parton level signatures change after hadronization and in the presence of initial state and final state
radiations. This changes the final state jet structure and the number of jets can increase or decrease due to these effects. 
In our analysis we tag a parton level tau as  $\tau$-jet via its hadronic decay with at least one charged track within $\Delta R \leq 0.1$ of the
candidate $\tau$-jet \cite{taujet}. On the other hand, we tag a jet as $b$-jet from the secondary vertex 
reconstruction with single $b$-jet tagging efficiency of $0.5$ \cite{btag}.

The dominant Standard Model backgrounds are $t\bar{t}+\rm{jets}$, $t\bar{t}Z$, $t\bar{t}W$.
The background events (except $t\bar{t}$) are generated using CalcHEP \cite{calchep} and PYTHIA \cite{pythia},
then hadronized via PYTHIA. We use FastJet \cite{fastjet} for jet reconstruction.
$t\bar{t}$ with associated QCD jets pose serious threat to the signal. The cross section of such process is so large that
even light jet faking a $\tau$-jet or a $b$-jet can reduce signal-to-background ratio.
%The soft QCD-jets along with $t\bar{t}$ can be very dengarous as the light jets coming from such radiation can be faked as $\tau$-jets. 
We estimate $t\bar{t}+\rm{jets}$ events using ALPGEN \cite{alpgen} where we use {\tt MLM} \cite{Hoche:2006ph} prescription 
to avoid double counting of events with jets coming from hard scattering (described by matrix element method) and from soft radiation
(described by parton shower models). We assume a mistagging efficiency $10^{-2}$ \cite{taumistag}
for a QCD jet to fake a $\tau$-jet.
%We consider a mistagging efficiency of $10^{-2}$ \cite{taumistag}
 %of a QCD jet that can fake a $\tau$-jet.}

In the figures the distribution of various variables are plotted at the production level without any selection cuts, in order to know their trends
and differences with respect to the main SM backgrounds (due to low statistics for $t\bar{t}$-jets, we have not included that in the plots).
These distributions guide us for the selection cuts leading to various final states.
Figure~\ref{njjpt} (left) describes the jet-multiplicity ($n_j$) distributions for the signal BP1 and the dominant SM backgrounds $t\bar{t}$,
$t\bar{t}Z$ and $t Z W^\pm$. We see that for the signal the jet-multiplicity $n_j$ peaks at 6. In the right panel of 
Fig.~\ref{njjpt} we plot the  $p^j_T$  distribution for the first and second $p_T$ ordered jets for the signal BP1 and for the dominant SM background $t\bar{t}$. 
From Fig.~\ref{njjpt} one can see that at least the first and second jets are somewhat harder for the signal than for the background.
 %$\gsim 50$ GeV.
  
%%%%%%%%%%%%%%%%%%%%%% p_t^tau and pt^l distributions %%%%%%%%%%%%%%%
\begin{figure}[t]
\begin{center}
\includegraphics[width=.33\linewidth, angle=-90]{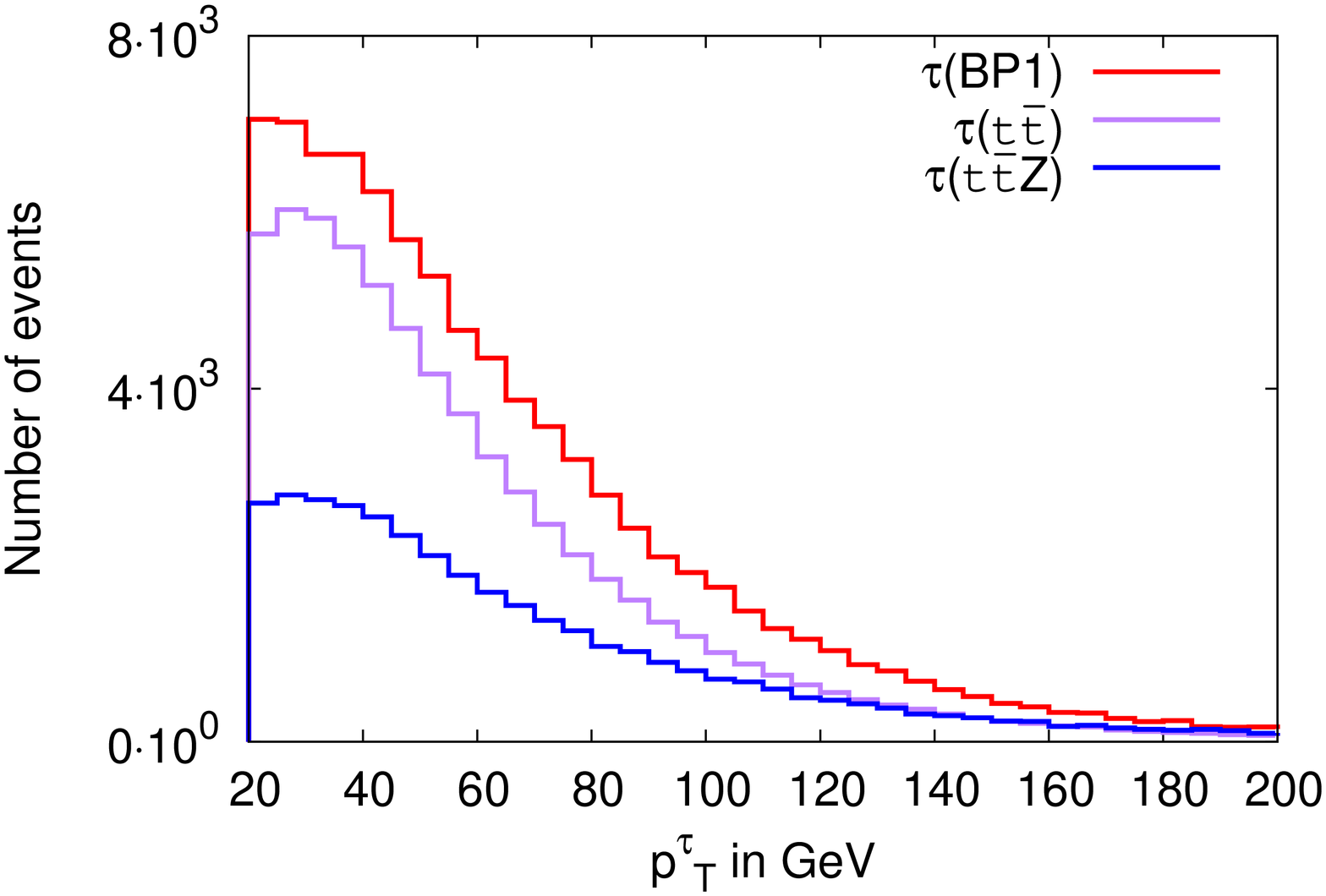}
\includegraphics[width=.33\linewidth, angle=-90]{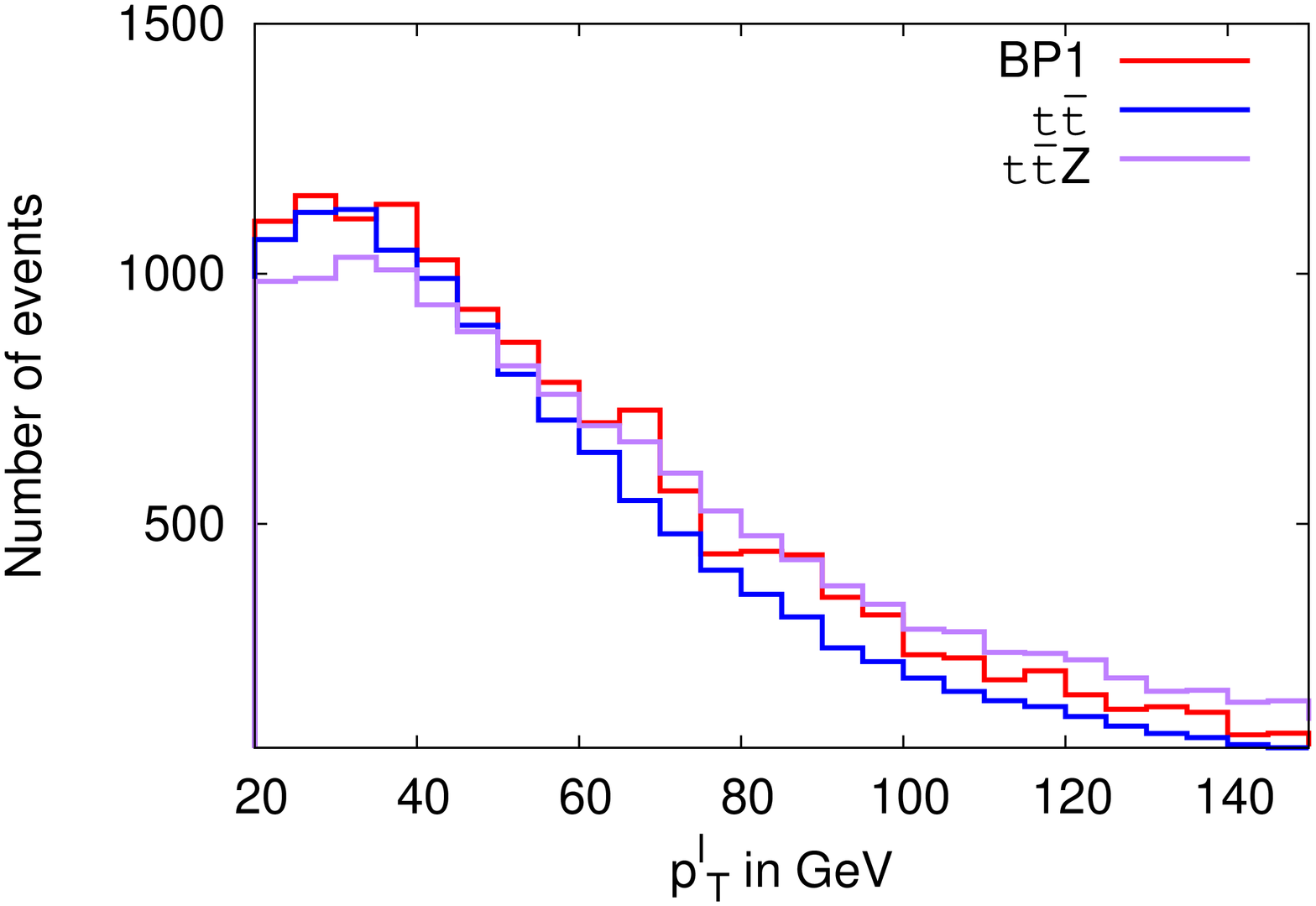}
\caption{$\tau$-jet $p_T$ distribution (left)  for the signal (BP1) and SM processes $t \bar{t}$, $t \bar{t} Z$.  
 $p_T$ distribution of the hardest lepton (right) for signal (BP1) and SM backgrounds $t\bar{t}$, $t\bar{t}Z$.}\label{taulpt}
\end{center}
\end{figure}
%%%%%%%%%%%%%%%%%%%%%%%%%%%%%%%%%%%%%%%%%%%%%%%%%%%%%%%%%%%%%%%%%%%

%%%%%%%%%%%%%%%%%%%%%% mtautau and mbb distributions %%%%%%%%%%%%%%%
\begin{figure}[b]
\begin{center}
\includegraphics[width=0.33\linewidth, angle=-90]{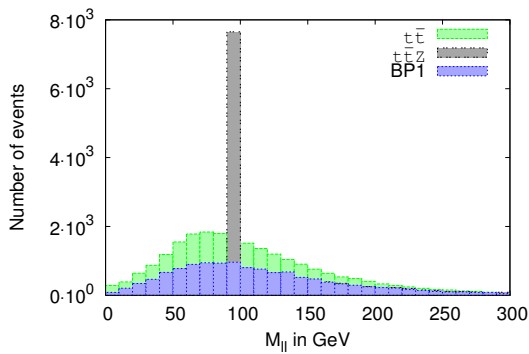}
\caption{ Lepton pair ($ee, \mu\mu$) invariant mass distribution for the signal (BP1) and dominant backgrounds $t\bar{t}$, $t\bar{t}Z$.}\label{mll}
\end{center}
\end{figure}
%%%%%%%%%%%%%%%%%%%%%%%%%%%%%%%%%%%%%%%%%%%%%%%%%%%%%%%%%%%%%%%%%%%
The left panel of Fig.~\ref{taulpt} shows the $p_{T}$ distribution of $\tau$-jet for the signal (BP1) and for SM backgrounds $t\bar{t}$ and $t\bar{t}Z$. 
We see that though the $\tau$s are coming from the hidden pseudoscalar decay for the signal, they are hard enough ($p_T \gsim$ 30 GeV) due to relatively heavier 
$a_1$ (around 70 GeV). Figure~\ref{taulpt} (right) shows the lepton $p_T$ distribution for the signal (BP1) and for the SM 
backgrounds $t\bar{t}$, $t\bar{t}Z$. The lepton $p_T$s are hard enough to be detected as hard lepton as they are coming from the decays of the gauge bosons. 
It can be seen from Fig.~\ref{mll} that the lepton pair coming from $Z$ mediated background like $t\bar{t}Z$ peaks 
around $m_Z$ in their invariant mass distribution which is not the case for the signal as they are coming from $W^\pm$ bosons. Thus we can use 
$|m_{\ell\ell} - m_Z|$ veto to kill the SM backgrounds coming from $Z$ boson.

\subsection{$1b + 2\tau + 2\ell +\etmiss$}

%%%%%%%%%%%%%%%%% 1b +2 \tau + 2l at 13 TeV %%%%%%%%%%%%%%%%%%%%  table 7
\begin{table}[t]
\begin{center}
\renewcommand{\arraystretch}{1.4}
\begin{tabular}{||c||c|c|c|c|c|c||c|c|c||}
\hline\hline
&\multicolumn{6}{|c||}{Benchmark}&\multicolumn{3}{|c||}{Backgrounds} \\
%\hline
\cline{2-10}
Final States/Cuts&\multicolumn{2}{|c|}{BP1} & \multicolumn{2}{|c|}{BP2}& \multicolumn{2}{|c||}{BP3} & & &\\
%\hline
\cline{2-7}
& $h^\pm t$ &$h^\pm bt$& $h^\pm t$ &$h^\pm bt$& $h^\pm t$ &$h^\pm bt$& $t\bar{t}+$jets &$tZW^\pm$& $t\bar{t}Z$\\
\hline\hline
%$n_j\leq 5$ $+2\ell$ & \multirow{2}{*}{19.89}& \multirow{2}{*}{41.94}&\multirow{2}{*}{100.56}&\multirow{2}{*}{78.37}&\multirow{2}{*}{245.96}&\multirow{2}{*}{84.94}&\multirow{2}{*}{13.1}&\multirow{2}{*}{49.66}\\
% &&&&&&&&\\
$n_j\leq 5$ $+ ~2\ell$ & 19.89 & 41.94 & 100.56 & 78.37 & 245.96 & 84.94 & 297.69 & 13.1 & 49.66 \\
$+|M_{\ell \ell}-M_Z|\geq 5$ GeV & 18.90 & 38.90 & 98.23 & 76.41 & 240.62 & 84.94 & 278.95 & 10.07 & 32.54 \\
$+|m_{\tau\tau}-M_Z|>10$ GeV & 18.90 & 37.07 & 93.55 & 74.45 & 213.88 & 84.94 & 155.95 & 7.05 & 23.98 \\
%$+m_{\tau\tau}<125.0$ GeV&40.13&&&&&&9.70&22.83\\
\hline
 Significance&\multicolumn{2}{|c|}{3.59}&\multicolumn{2}{|c|}{8.92}&\multicolumn{2}{|c||}{13.56}&\multicolumn{3}{|c||}{--}\\
\hline
%\multirow{3}{*}{$p_1:|m_{\tau\tau}-m_{a_1}|\leq 10\,$GeV}&\multicolumn{2}{|c|}{}&\multicolumn{2}{|c|}{}&\multicolumn{2}{|c||}{} & 0.30 & 1.20 \\
%&\multicolumn{2}{|c|}{1.67}&\multicolumn{2}{|c|}{7.57}&\multicolumn{2}{|c||}{14.22} & 0.40 & 1.20\\
%&\multicolumn{2}{|c|}{}&\multicolumn{2}{|c|}{}&\multicolumn{2}{|c||}{} & 0.40 & 1.37\\
%\hline
%\hline
% Significance&\multicolumn{2}{|c|}{0.94}&\multicolumn{2}{|c|}{2.50}&\multicolumn{2}{|c||}{3.56}&\multicolumn{2}{|c||}{--}\\
%\hline
\hline
\end{tabular}
\caption{The number of events for $n_j\leq 5$ (includes $1b$-jet+ $2\tau$-jet) + $2\ell$ final state at 1000 fb$^{-1}$ of
luminosity at the LHC for center of mass energy (ECM) of 13 TeV. }\label{b2tau2l13}
\end{center}
\end{table}

%%%%%%%%%%%%%%%%%%%%%%
%For relatively heavier charged Higgs with mass $m_{h^\pm}>m_t$, the main 
%production process is $b$, gluon fusion, i.e., $bg\to t h^\pm$ (see Figure~\ref{cross}).
%The charged Higgs boson thus produced will decay to the light pseudoscalar $a_1$
%and  $W^\pm$ boson. Being light, $m_{a_1}< 100$ GeV, the pseudoscalar dominantly 
%decays to $b$ and $\tau$ pairs. We investigate both the decay modes in this study. 
First, we consider the pseudoscalar decay to a pair of $\tau$ jets in association with leptons coming from both the $W^\pm$.
The final state, thus, becomes $1b+2\tau +2\ell +\etmiss$. This is relatively clean when compared with SM backgrounds.
The $b$ and $\tau$ tagging reduce the dominant di-lepton backgrounds coming from the  gauge boson pair in association with jets.
The requirement of lower number of jets $\leq 5$ and a veto on di-lepton invariant
mass around $M_Z$ further reduce such backgrounds.  Nevertheless, we see that there are
events coming from  $t\bar{t}+\text{jets}$, $tZW^\pm$, $t\bar{t}Z$.

In Tables~\ref{b2tau2l13} and \ref{b2tau2l14}, we present the number of events coming from the signal for the three benchmark points and the SM backgrounds at an integrated luminosity 
of 1000 fb$^{-1}$ at 13 TeV and 14 TeV center of mass energy at the LHC, respectively. We can see that $b$-jet and $\tau$-jet invariant mass veto cuts around $M_Z$ 
help reduce the SM backgrounds. At this stage benchmark points BP2 and BP3 cross $5\sigma$ signal significance 
with BP3 being the highest for both cases. This shows that as early as 136 (122) fb$^{-1}$ some parameter points can be probed at the LHC with ECM
of 13 (14) TeV. For BP1 and BP2 the signal significances are 3.59 (4.03) and 
8.92 (8.65) respectively for 13 (14) at the LHC with 1000 fb$^{-1}$ of integrated luminosity.

\begin{figure}[b]
\begin{center}
\includegraphics[width=.43\linewidth, angle=-90]{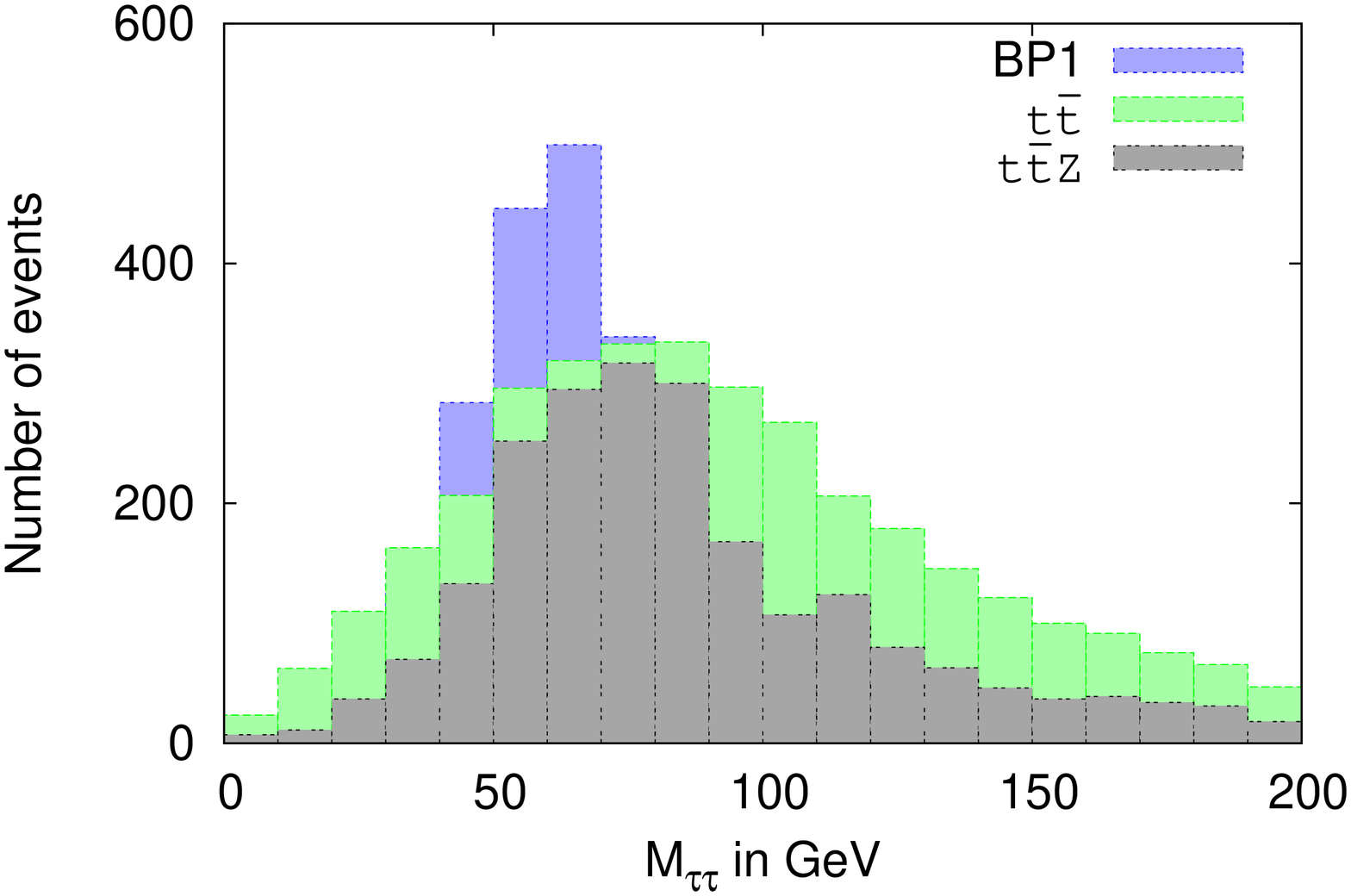}
\caption{$\tau_{\rm jet}$ pair invariant mass distribution for the signal (BP1) 
and dominant backgrounds $t\bar{t}$, $t\bar{t}Z$.}\label{mtautau}
\end{center}
\end{figure}

We have defined $\tau$-jet via its hadronic decay with one prong charged track. A 
light pseudoscalar when decaying into tau pairs can give rise to two hadronic $\tau$-jets.
Their invariant mass is described as $m_{\tau\tau}$ and the distribution is shown in Figure~\ref{mtautau}.
%Unlike the $b$-jets coming from the top decays, the $\tau$-jets coming from the light pseudoscalar ($a_1$) are rather soft and their invariant mass peak
%around $m_{a_1}$ as shown in Figure~\ref{mtautau}. 

%%%%%%%%%%%%%%%%% b + 2 \tau+ 2l at 14 TeV %%%%%%%%%%%%%%%%%%%%  table 8
\begin{table}
\begin{center}
\renewcommand{\arraystretch}{1.4}
\begin{tabular}{||c||c|c|c|c|c|c||c|c|c||}
\hline\hline
&\multicolumn{6}{|c||}{Benchmark}&\multicolumn{3}{|c||}{Backgrounds} \\
%\hline
\cline{2-10}
Final States/Cuts&\multicolumn{2}{|c|}{BP1} & \multicolumn{2}{|c|}{BP2}& \multicolumn{2}{|c||}{BP3} & & &\\
%\hline
\cline{2-7}
& $h^\pm t$ &$h^\pm bt$& $h^\pm t$ &$h^\pm bt$& $h^\pm t$ &$h^\pm bt$ & $t\bar{t}+$jets &$tZW^\pm$& $t\bar{t}Z$\\
\hline\hline
$n_j\leq 5$ $+~ 2\ell $ & 26.67 & 55.76 & 103.18 & 79.65 & 345.72 & 52.42 & 320.23 & 14.54 & 51.88\\
% &&&&&&&&\\
$+|M_{\ell \ell}-M_Z|\geq 5$ GeV & 20.32 & 48.97 & 103.18 & 74.82 & 300.06 & 52.42 & 299.06 & 13.57 & 41.51\\
$+|m_{\tau\tau}-M_Z|>10$ GeV  & 19.05 & 47.47 & 94.6 & 72.41 & 280.49 & 52.42 &165.12 & 9.70 & 31.13\\
%$+m_{\tau\tau}<125.0$ GeV&40.13&&&&&&9.70&22.83\\
\hline
 Significance&\multicolumn{2}{|c|}{4.03}&\multicolumn{2}{|c|}{8.65}&\multicolumn{2}{|c||}{14.34}&\multicolumn{3}{|c||}{--}\\
\hline
%\multirow{3}{*}{$p_1:|m_{\tau\tau}-m_{a_1}|\leq 10\,$GeV}&\multicolumn{2}{|c|}{}&\multicolumn{2}{|c|}{}&\multicolumn{2}{|c||}{}&0.48&0.83\\
%&\multicolumn{2}{|c|}{1.86}&\multicolumn{2}{|c|}{6.37}&\multicolumn{2}{|c||}{10.43} & 0.68 & 1.25\\
%&\multicolumn{2}{|c|}{}&\multicolumn{2}{|c|}{}&\multicolumn{2}{|c||}{}&0.48&1.04\\
%\hline
%\hline
% Significance&\multicolumn{2}{|c|}{1.04}&\multicolumn{2}{|c|}{2.21}&\multicolumn{2}{|c||}{3.02}&\multicolumn{2}{|c||}{--}\\
%\hline
\hline
\end{tabular}
\caption{The number of events for $n_j\leq 5$ (includes $1b$-jet+ $2\tau$-jet) + $2\ell$ final state at 1000 fb$^{-1}$ of 
luminosity at the LHC for center of mass energy (ECM) of 14 TeV.}\label{b2tau2l14}
\end{center}
\end{table}

%%%%%%%%%%%%%%%%%%%%%%%%%%%%%%%%%%
%%%%%%%%%%%%%%%%%%%%%% mtautau distributions %%%%%%%%%%%%%%%
%%%%%%%%%%%%%%%%%%%%%%%%%%%%%%%%%%%%%%%%%%%%%%%%%%%%%%%%%%%%%%%%%%%

\subsection{$1b + 2\tau + 2j + 1\ell +\etmiss$}

%%%%%%%%%%%%%%%%%%%%%%%   ($1b$-jet+ $2\tau$-jet) + $1\ell  @ 13 TeV    table 9 %%%%%%%%%%%%%%%%%
\begin{table}[t]
\begin{center}
\hspace*{-2.0cm}
\renewcommand{\arraystretch}{1.4}
\begin{tabular}{||c||c|c|c|c|c|c||c|c|c||}
\hline\hline
&\multicolumn{6}{|c||}{Benchmark}&\multicolumn{3}{|c||}{Backgrounds} \\
%\hline
\cline{2-10}
Final States/Cuts&\multicolumn{2}{|c|}{BP1} & \multicolumn{2}{|c|}{BP2}& \multicolumn{2}{|c||}{BP3} & & &\\
%\hline
\cline{2-7}
& $h^\pm t$ &$h^\pm bt$& $h^\pm t$ &$h^\pm bt$& $h^\pm t$ &$h^\pm bt$ & $t\bar{t}+$jets & $tZW^\pm$ & $t\bar{t}Z$ \\
\hline
\hline
%$n_j\leq 6$($1b$-jet+ $2\tau$-jets)&\multirow{2}{*}{163.10}& \multirow{2}{*}{495.34}&\multirow{2}{*}{1014.99}&\multirow{2}{*}{854.21}&\multirow{2}{*}{2571.91}&\multirow{2}{*}{1104.25}&\multirow{2}{*}{81.59}&\multirow{2}{*}{371.62}\\
% &&&&&&&&\\
$n_j\leq 6$ & 163.10 &{495.34}&{1014.99}& 854.21&2571.91&1104.25& 2524.62 & 81.59 & 371.62\\
$+|M_{j j}-M_W|\leq 10$ GeV & 108.40 & 352.51 & 720.32 & 624.98 & 1919.57 & 962.68 & 1783.25 & 64.47 & 309.97\\
$+|m_{\tau\tau}-M_Z|>10$ GeV & 97.46 & 318.48 & 643.14 & 548.58 & 1716.39 & 877.73 & 372.78 & 51.37 & 232.90\\
$+m_{\tau\tau}<125.0$ GeV & 92.49 & 288.09 & 605.72 & 505.47 & 1641.53 & 849.42 & 372.78 & 49.36 & 195.23 \\
\hline
 Significance&\multicolumn{2}{|c|}{12.05}&\multicolumn{2}{|c|}{26.73}&\multicolumn{2}{|c||}{44.68}&\multicolumn{3}{|c||}{--}\\
\hline
\hline
%\multirow{3}{*}{$p_1:|m_{\tau\tau}-m_{a_1}|\leq 10\,$GeV}&\multicolumn{2}{|c|}{}&\multicolumn{2}{|c|}{}&\multicolumn{2}{|c||}{} & 2.72 & 11.30\\
%&\multicolumn{2}{|c|}{13.57}&\multicolumn{2}{|c|}{40.55}&\multicolumn{2}{|c||}{85.05} & 2.12 & 10.96 \\
%&\multicolumn{2}{|c|}{}&\multicolumn{2}{|c|}{}&\multicolumn{2}{|c||}{} & 2.52 & 10.28 \\
%\hline
%\hline
% Significance&\multicolumn{2}{|c|}{2.58}&\multicolumn{2}{|c|}{5.54}&\multicolumn{2}{|c||}{8.60}&\multicolumn{2}{|c||}{--}\\
%\hline
%\hline
\end{tabular}
\caption{The number of events for $n_j\leq 6$ (includes $1b$-jet+ $2\tau$-jet) + $1\ell$  final state at 1000 fb$^{-1}$ of
luminosity at the LHC for center of mass energy (ECM) of 13 TeV.}\label{b2taul13}
\end{center}
\end{table}
%%%%%%%%%%%%%%%%%%%%%%%%%%%%%%%%%%%%%
%

In this part we consider the case when one of the $W^\pm$ decays hadronically. The advantage is the enhancement in signal number by a
combinatoric factor of two as the other $W^\pm$ still decays to leptons. 
Both the $\tau$ and $b$ tagging keep the SM backgrounds in control. Like in the previous case,
$t\bar{t}+\text{jets}$, $tZW^\pm$, and $t\bar{t}Z$ are the irreducible backgrounds. 

Table~\ref{b2taul13} and \ref{b2taul14} present the number of events for the benchmark points and the dominant SM 
backgrounds at an integrated luminosity of 1000 fb$^{-1}$ for 13 and 14 TeV center of mass energies at the LHC. For the
final state we demand $n_j\leq 6$ (which includes $1b$-jet+ $2\tau$-jets). The rest of the jets can come from ISR, FSR
or showering. Any two jets from the remaining three jets which are not tagged as $b$ or $\tau$-jets are required to have their invariant mass 
within 10 GeV of $M_W$, which reduces the combinatorial backgrounds. The requirement of ditau invariant mass
outside 10 GeV of the $Z$ boson mass reduces $t\bar{t}+$jets events severely. Finally we demand the $\tau$-jet pair invariant mass to be within 125 GeV
as we are looking for a light pseudoscalar which is lighter than the 125 GeV Higgs (but greater than half of it).

All the points cross $5\sigma$ signal significance for both 13 and 14 TeV energy at the LHC with the
highest for BP3 of about $45 \sigma $ and $51\sigma$, respectively. This shows that with a very early data, 
around 10 fb$^{-1}$ of integrated luminosity we can achieve $5\sigma$ significance at the LHC.

%As before, we are interested in the invariant mass peak resolution in the $b \bar{b}$ mode for all the  benchmark points.
%We achieve 2.96$\sigma$, 8.1$\sigma$, 6.54$\sigma$ significance in BP1, BP2, BP3 respectively at 1000 fb $^{-1}$ of
%integrated luminosity at 14 TeV LHC. 

%BP2 and BP3 even cross more than $5\sigma$ reach.

%%%%%%%%%%%%%%%%%%%%%%%        %%%%%%%%%%%%%%%%%  table -10 
\begin{table}
\begin{center}
\hspace*{-2.0cm}
\renewcommand{\arraystretch}{1.4}
\begin{tabular}{||c||c|c|c|c|c|c||c|c|c||}
\hline\hline
&\multicolumn{6}{|c||}{Benchmark}&\multicolumn{3}{|c||}{Backgrounds} \\
%\hline
\cline{2-10}
Final States/Cuts&\multicolumn{2}{|c|}{BP1} & \multicolumn{2}{|c|}{BP2}& \multicolumn{2}{|c||}{BP3}&&& \\
%\hline
\cline{2-7}
& $h^\pm t$ &$h^\pm bt$& $h^\pm t$ &$h^\pm bt$& $h^\pm t$ &$h^\pm bt$& $t\bar{t}+$jets & $tZW^\pm$& $t\bar{t}Z$ \\
\hline\hline
%$n_j\leq 6$ &\multirow{2}{*}{200.66}& \multirow{2}{*}{602.01}&\multirow{2}{*}{1252.48}&\multirow{2}{*}{1025.81}&\multirow{2}{*}{2994.10}&\multirow{2}{*}{2149.26}&\multirow{2}{*}{93.05}&\multirow{2}{*}{369.43}\\
% &&&&&&&&\\
$n_j\leq 6$ & {200.66}& {602.01} & {1252.48} & {1025.81} & {2994.10} & {2149.26} & 2978.86& 93.05 & {369.43}\\
$+|M_{j j}-M_W|\leq 10$ GeV & 153.67 & 437.01 & 879.89 & 731.34 & 2061.30 & 1834.74 & 2113.33 & 67.85 & 286.42\\
$+|m_{\tau\tau}-M_Z|>10$ GeV & 135.89 & 394.06 & 773.84 & 661.34 & 1846.04 & 1625.05  & 414.82 & 49.43 & 203.40\\
$+m_{\tau\tau}<125.0$ GeV & 115.57 & 366.94 & 742.31 & 605.83& 1806.90 & 1362.95 & 414.82 & 45.56 & 172.26 \\
\hline
 Significance&\multicolumn{2}{|c|}{14.45}&\multicolumn{2}{|c|}{30.29}&\multicolumn{2}{|c||}{51.40}&\multicolumn{3}{|c||}{--}\\
\hline
%\multirow{3}{*}{$p_1:|m_{\tau\tau}-m_{a_1}|\leq 10\,$GeV}&\multicolumn{2}{|c|}{}&\multicolumn{2}{|c|}{}&\multicolumn{2}{|c||}{}&2.52&7.89\\
%&\multicolumn{2}{|c|}{16.64}&\multicolumn{2}{|c|}{55.10}&\multicolumn{2}{|c||}{126.09}&2.33&8.30\\
%&\multicolumn{2}{|c|}{}&\multicolumn{2}{|c|}{}&\multicolumn{2}{|c||}{}&2.81&8.71\\
%\hline
%\hline
% Significance&\multicolumn{2}{|c|}{3.20}&\multicolumn{2}{|c|}{6.79}&\multicolumn{2}{|c||}{10.74}&\multicolumn{2}{|c||}{--}\\
%\hline
\hline
\end{tabular}
\caption{The number of events for $n_j\leq 6$ (includes $1b$-jet+ $2\tau$-jet) + $1\ell$  final state at 1000 fb$^{-1}$ of
luminosity at the LHC for center of mass energy (ECM) of 14 TeV.}\label{b2taul14}
\end{center}
\end{table}

%%%%%%%%%%%%%%%%%%%%%%%%%%%%%%%%%%

%%%%%%%%%%%%%%%%%%%%%%%%%%%%%%%%%%

\subsection{$3b + 2\ell + \etmiss$}
%

%%%%%%%%%%%%%% 3b +2l at 13 TeV %%%%%%%%%%%%%%%%%%%%%%%% -- TABLE 11
\begin{table}[t]
\begin{center}
\hspace*{-2.5cm}
\renewcommand{\arraystretch}{1.0}
\begin{tabular}{||c||c|c|c|c|c|c||c|c|c|c||}
\hline
&\multicolumn{6}{|c||}{Benchmark}&\multicolumn{4}{|c||}{Backgrounds}\\
\cline{2-11}
Final states/Cuts&\multicolumn{2}{|c|}{BP1} & \multicolumn{2}{|c|}{BP2}&\multicolumn{2}{|c||}{BP3} &  &  &  &\\
\cline{2-7}
& $h^\pm t$ &$h^\pm bt$& $h^\pm t$ & $h^\pm bt$ & $h^\pm t$ & $h^\pm bt$ & $t\bar{t}$ &  $t\bar{t}Z$ & $tbW$ & $tZW^\pm$\\
\hline\hline
%$n_j\leq 5$ ($3b$-jet) + $2\ell$ & 261.46 & 145.87 & 496.50 & 454.53 & 623.40 & 425.92 & 4165.80 & 39.05 & 8156.15 & 8.16 \\
$n_j\leq 5$ + $2\ell$ & 261.46 & 145.87 & 496.50 & 454.53 & 623.40 & 425.92 & 4165.80 & 39.05 & 8156.15 & 8.16 \\

$+|m_{\ell\ell}-M_Z|>5$ GeV & 243.96 & 136.14 & 463.29 & 423.58 & 567.09 & 362.03 & 3812.48 & 34.76 & 7422.68 & 7.66 \\

$+|m_{bb}-M_{Z}|>10$ GeV & 149.28 & 94.51 & 297.01 & 273.11 & 394.15 & 191.66 & 2538.92 & 14.04 & 5104.93 & 3.22 \\

$+m_{bb}<125$ GeV & 133.27 & 85.39 &  282.75 & 245.88 & 378.06 & 170.37 & 1774.78 & 12.16 & 3520.64 & 2.82\\

$+p_T^{bj_{2,3}}<100$GeV & 121.33 & 74.45 & 260.06 & 221.0 & 357.95 & 170.37 & 1528.28 & 9.59 & 3168.58 & 1.71\\
\hline
Significance&\multicolumn{2}{|c|}{2.8}&\multicolumn{2}{|c|}{6.68}&\multicolumn{2}{|c||}{7.3}&\multicolumn{4}{|c||}{--}\\
\hline
\multirow{3}{*}{$p_1:|m_{bb}-m_{a_1}|\leq 10\,$GeV}&\multicolumn{2}{|c|}{}&\multicolumn{2}{|c|}{}&\multicolumn{2}{|c|}{} & 534.08 & 5.48 & 1202.89 & 1.21 \\
&\multicolumn{2}{|c|}{97.87}&\multicolumn{2}{|c|}{290.88}&\multicolumn{2}{|c|}{239.67 } & 591.59 & 4.80 & 1202.89 & 1.11\\
&\multicolumn{2}{|c|}{}&\multicolumn{2}{|c|}{}&\multicolumn{2}{|c|}{} & 599.81 &  5.82 & 1408.26 & 1.41 \\
\hline
 Significance&\multicolumn{2}{|c|}{2.28}&\multicolumn{2}{|c|}{6.36}&\multicolumn{2}{|c|}{5.04}&\multicolumn{4}{|c|}{--}\\
\hline
\hline
\end{tabular}
\caption{The number of events for $n_j\leq 5$ (includes $3b$-jet) + $2\ell$  final state at 100 fb$^{-1}$ of
luminosity at the LHC for center of mass energy (ECM) of 13 TeV. }\label{3b2l13}
\end{center}
\end{table}
%%%%%%%%%%%%%%%%%%%%%%%%%%%%%%%%%%%%%%%%%%%%%%
%%%%%%%%%%%%%%%%%%%%%% mbb distributions %%%%%%%%%%%%%%%
\begin{figure}[t]
\begin{center}
\includegraphics[width=.33\linewidth, angle=-90]{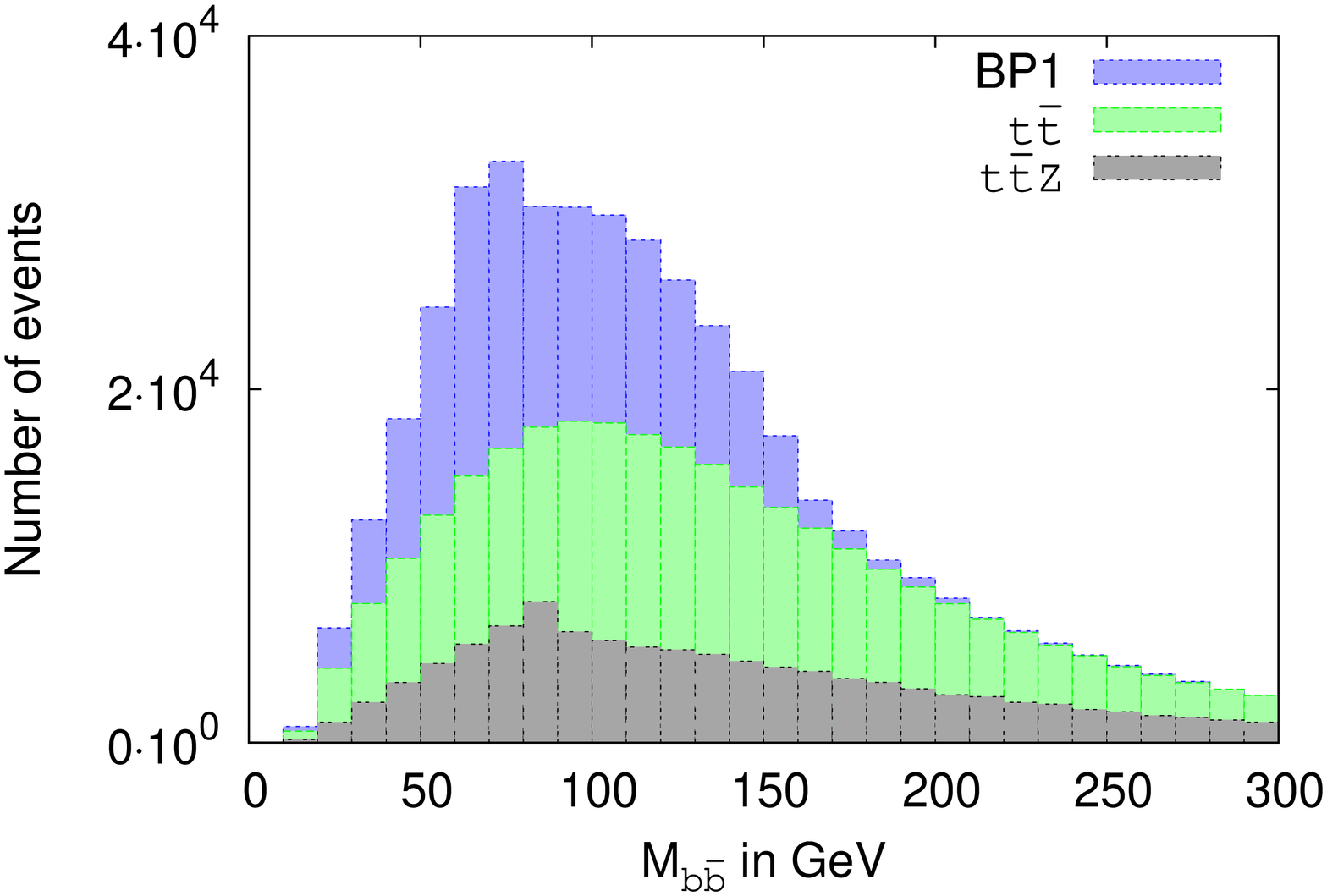}
\caption{$b$-jet pair invariant mass distribution for the signal (BP1) and dominant backgrounds $t\bar{t}$, $t\bar{t}Z$.}\label{mbb}
\end{center}
\end{figure}
%%%%%%%%%%%%%%%%%%%%%%%%%%%%%%%%%%%%%%%%%%%%%%%%%%%%%%%%%%%%%%%%%%%

Finally, we consider the case where the light pseudoscalar decays to a $b$ pair.
This gives rise to a final state which constitutes of $3b+2\ell +\etmiss$ with both the $W^{\pm}$s decaying
leptonically. However, at the jet level we demand $n_j\leq 5$ (which includes $3 b$-jets).

Table~\ref{3b2l13} and \ref{3b2l14} present the number of events for the three benchmark points and the SM backgrounds. 
Removal of the $\tau$-tagging from the previous cases increases the SM background contribution. This includes 
$t\bar{t}$, $t\bar{t}Z$, $tbW$ and $tZW^\pm$. To reduce these contributions we apply lepton pair invariant mass
veto and $b$-jet pair invariant mass veto around the $Z$ boson mass. As in the previous case for the $\tau$-pair, we 
demand $b$-jet pair invariant mass to lie within 125 GeV to confirm that they can come only from the light state below 125 GeV.
However, the behaviour of $tbW$ background for this final state for 13 and 14 TeV energies
is not very intuitive. We can see from Table~\ref{3b2l13} and \ref{3b2l14} that the 13 TeV numbers
are greater than 14 TeV for $tbW$ process. We check via detail simulation that hadronic activity around leptons makes lepton isolation difficult for
$tbW$. It is understandable that jet activity around a lepton is much enhanced as the center-of-mass energy increases from 13 TeV to 14 TeV at the LHC.
This makes the number of isolated leptonic events for ECM $= 14$ TeV much smaller than in case of 13 TeV.
One may think that a similar observation should hold true for $t \bar t$ as well.
We stress that the choice of our final state makes interesting impact in this case. A generic final state without
any QCD radiation from $tbW$ gives $2b + 2l + \etmiss$ assuming both the $W$s decay leptonically. 
In this case, 14 TeV numbers are greater than 13 TeV, as expected. However, in our case, the extra $b$-jet (as we demand $3b$) has to come from QCD radiation
({\it viz}, radiation from final state top or bottom) faking as $b$-jet. Hence, we are bound to take resort of
extra QCD jets. Gluon emission from top does not affect much, but radiation coming from $b$ quark is more collinear for 
14 TeV than for 13 TeV center-of-mass energy. Hence, jet-jet isolation 
criteria becomes too tight in case of 14 TeV. Thus number of events qualifying the cuts for 14 TeV
become smaller than the 13 TeV numbers. On the other hand, in case of two on-shell top quarks, QCD radiation
coming from one of the top quarks is enough. % and does not have to depend on radiation from bottom quark.
QCD gluon emission is more from on-shell top than from final state bottom. Hence, the same argument does not hold for
$t \bar t$ scenario, as the $b$s (coming from top decay) and extra jet (coming as a radiation from top) are well separated to pass isolation criteria
for both the center-of-mass energies.

Out of 3 $b$-jets two are coming from the light pseudoscalar in the case of the signal so we further require the $p_T$ of the second and third $p_T$ order
$b$-jets to be less than 100 GeV. The signal significance at this stage are 2.8 (3.96), 6.68 (8.97) and 7.30 (12.11)$\sigma$
for BP1, BP2 and BP3, respectively, for ECM 13 (14) at the LHC.

%Unlike previous two cases here we have relatively larger events around the pseudoscalar mass peak.
Figure~\ref{mbb} shows $b$-jet pair invariant mass distribution for the BP1 and for the dominant backgrounds $t\bar{t}$ and 
$t\bar{t}Z$, respectively. In addition to the standard cuts as shown in Table~\ref{3b2l13} and \ref{3b2l14}, we can also use $b \bar{b}$
invariant mass cut (which peaks around the pseudoscalar mass) to achieve fairly good significance. 
This also gives a hint of the region where pseudoscalar mass may lie. Selection of events around 10 GeV of the $b\bar{b}$ invariant mass peak provides 
2.28 (2.96), 6.36 (8.10) and 5.04 (8.19)$\sigma$ signal significances for BP1, BP2 and BP3 respectively for 13 (14) TeV at the LHC
%We probe the light pseudoscalar mass peak for the benchmark points and the signal significance for such mass resolutions are 2.28 (2.96), 6.36 (8.10) 
%and 5.04 (8.19)$\sigma$ for BP1, BP2 and BP3 respectively for 13 (14) TeV at the LHC. 
The $b\bar{b}$ peak is rather broad in Figure~\ref{mbb}, mainly because of combinatoric factor. If we increase the selection window
to $\pm 15$ ($20$) GeV around the light pseudoscalar mass peak in $b\bar{b}$ invariant mass
distribution, the signal significances enhance upto 15\% (23\%) depending on the benchmark point.

%%%%%%%%%%%%%% 3b +2l at 14 TeV %%%%%%%%%%%%%%%%%%%%%%%% -- TABLE 12
\begin{table}
\begin{center}
\hspace*{-2.5cm}
\renewcommand{\arraystretch}{1.1}
\begin{tabular}{||c||c|c|c|c|c|c||c|c|c|c||}
\hline
&\multicolumn{6}{|c||}{Benchmark}&\multicolumn{4}{|c||}{Backgrounds}\\
\cline{2-11}
Final states/Cuts&\multicolumn{2}{|c|}{BP1} & \multicolumn{2}{|c|}{BP2}&\multicolumn{2}{|c||}{BP3} &  &  &  &\\
\cline{2-7}
& $h^\pm t$ &$h^\pm bt$& $h^\pm t$ & $h^\pm bt$ & $h^\pm t$ & $h^\pm bt$ & $t\bar{t}$ &  $t\bar{t}Z$ & $tbW$ & $tZW^\pm$\\
%&&&&&&&&&&&+jets\\
\hline\hline
$n_j\leq 5$ + $2\ell$ & 313.82 & 180.30 & 564.33 & 546.45 & 804.63 & 670.30 & 5372.78 & 54.59 & 3588.78 & 9.89\\

$+|m_{\ell\ell}-M_Z|>5$  & 292.99 & 166.74 & 523.06 & 509.04 & 726.13 & 630.90 & 4920.12 & 47.53 & 3135.83 & 8.43\\

$+|m_{bb}-M_{Z}|>10$  & 188.470 & 114.60 & 343.93 & 330.43 & 520.06 & 473.10 & 3158.72 & 21.58 & 1986.03 & 4.46\\

$+m_{bb}<125$  & 167.26 & 104.13 &  328.17 & 299.29 & 466.10 & 394.30 & 2292.78 & 16.19 & 1742.13 & 4.36\\

$+p_T^{bj_{2,3}}<100$ & 150.88 & 92.45 & 304.09 & 270.33 & 441.56 & 354.80 & 1977.89 & 13.08 & 1533.07 & 3.78\\
\hline
Significance&\multicolumn{2}{|c|}{3.96}&\multicolumn{2}{|c|}{8.97}&\multicolumn{2}{|c||}{12.11}&\multicolumn{4}{|c||}{--}\\
\hline
\multirow{3}{*}{$p_1:|m_{bb}-m_{a_1}|\leq 10\,$GeV}&\multicolumn{2}{|c|}{}&\multicolumn{2}{|c|}{}&\multicolumn{2}{|c|}{} & 767.54 & 7.06 & 731.69  & 2.62\\
&\multicolumn{2}{|c|}{119.46}&\multicolumn{2}{|c|}{347.50}&\multicolumn{2}{|c|}{358.52 }& 787.22 & 6.23 & 696.85 & 2.42 \\
&\multicolumn{2}{|c|}{}&\multicolumn{2}{|c|}{}&\multicolumn{2}{|c|}{}& 816.74 & 7.89 &  731.69  & 2.81\\
\hline
Significance&\multicolumn{2}{|c|}{2.96}&\multicolumn{2}{|c|}{8.1}&\multicolumn{2}{|c|}{8.19}&\multicolumn{4}{|c||}{--} \\
\hline
\hline
\end{tabular}
\caption{The number of events for $n_j\leq 5$ (includes $3b$-jet) + $2\ell$  final state at 100 fb$^{-1}$ of 
luminosity at the LHC for center of mass energy (ECM) of 14 TeV.}\label{3b2l14}
\end{center}
\end{table}

%%%%%%%%%%%%%%%%%%%%%%%%%%%%%%%%%%%%%%%%%%%%%%

\section{Discussion and conclusions}\label{conc}

In this paper we have considered the possibility of a hidden pseudoscalar ($\leq 100$ GeV) and a relatively light charged Higgs
(just above $m_t$) decaying into it in the NMSSM framework. Such hidden pseudoscalar is required to have an appropriate
singlet-doublet mixing in order to evade LEP bound as well as to have coupling with charged Higgs. This decay mode of the charged
Higgs $h^{\pm } \rightarrow a_1 W^{\pm}$ has not been searched by ATLAS \cite{ChATLAS} or CMS \cite{ChCMS} at the LHC, 
where finding a parameter region with a substantial branching ratio is difficult to get given the complicated parameter dependence. 
We have taken up a detailed collider analysis on this mode to highlight that this mode can be useful in exotic searches at the LHC.

%This decay mode of charged 
%Higgs $h^{\pm } \rightarrow a_1 W^{\pm}$ is non-standard in the sense that difficult to obtain a healthy branching
%ratio given the complicated parameter dependence. This channel has not been probed so far by
%ATLAS \cite{ChATLAS} and CMS \cite{ChCMS}. Studies on this mode with detail collider analysis 
%are not abound. Hence, we take up this study to highlight this mode which can be helpful in exotic 
%searches at the LHC.

First, we scanned a seven dimensional parameter space using the publicly available code NMSSMTools v4.7.0. We demanded the lightest
CP even Higgs to have mass around 125 GeV and also to satisfy the other experimental results from the LHC. We found a suitable parameter region
with a light pseudoscalar and also large branching fraction $h^{\pm } \rightarrow a_1 W^{\pm}$. We selected three benchmark points. $\tan \beta$ is a crucial 
parameter in the Higgs sector. We saw that in different $\tan \beta$ regions
(low, moderate and high), the charged Higgs can be just heavier than the top quark and simultaneously have a large branching ratio to $a_1$.

%low, moderate and high, charged Higgs mass can be just above than top quark and simultaneously have large branching to $a_1$.

Next, we discussed the main production processes of the charged Higgs boson at the LHC. The cross section for the associated
production with top quark, {\it i.e.} $pp \to t h^\pm$ and $pp \to b t h^\pm$, is larger than for  the charged Higgs pair production.
Like MSSM, NMSSM has only one physical charged Higgs boson $h^\pm$ and it is doublet type as singlet does not contribute
to charged Higgs boson. The other production channels ({\it e.g.} charged Higgs in association with a gauge boson
or SM-like Higgs) have typically smaller cross section.

The presence of a light pseudoscalar gives $b$- or $\tau$-rich final state which helps to avoid the SM backgrounds. 
We investigated the $1b + 2\tau + 2\ell +\ptmiss$, $1b + 2\tau + 2j + 1\ell + \ptmiss$ and $3b + 2\ell + \ptmiss$
final states resulting from $W^{\pm}$ decay modes. A detailed cut-based analysis was performed in order to find a reasonably positive
result in favour of our signal. We found that such  scenarios can be probed with the data of as little 
as $\sim 10$ fb$^{-1}$ of integrated luminosity at the LHC with 13 TeV and 14 TeV center-of-mass energy.  
%Even the $\tau$-jet pair invariant mass distribution can resolve the mass peak for the 
%hidden pseudoscalar in this case. 

Hidden scalars are still possible with the recent data from LHC, especially in the context of 
triplet-singlet extended Higgs sectors with $Z_3$ symmetries \cite{hiddenHE}. In MSSM the heavier Higgs bosons
($h_2, a, h^\pm$) are almost degenerate  which rules out the possibility of $h^\pm \rightarrow a W^\pm$,
where $a$ is the only massive pseudoscalar. In the case of NMSSM such hidden scalar is still allowed by LHC data
and its presence prompts the decay  $h^\pm \to a_1 W^\pm$ which is not 
possible in the CP-conserving MSSM. 
In CP-violating MSSM it is possible to find a very light mostly CP-odd hidden scalar,
and charged Higgs can indeed decay to $h_1 W^\pm$ \cite{CPVMSSM}. The triplet extended scenarios have also
charged Higgs along with pseudoscalars, and can have new features, for {\it e.g.}, the $Y=0$ triplet-type charged Higgs does not couple to $tb$ or $\tau \nu$
\cite{tripch}. Distinguishing such charged Higgs bosons of different representations may also be possible at the LHC \cite{Bandyopadhyay:2015ifm}.

%The case of $Y=\pm 2$ hypercharged triplet and singlet extended 
%supersymmetric scenario with $Z_3$ symmetry can also have a light pseudoscalar called $R$-axion.
%This scenario constitutes of more charged Higgs including a doubly charged Higgs boson $h^{\pm\pm}$ \cite{agashe}.

Finding a charged Higgs boson will be a proof of the existence of at least another $SU(2)_L$ doublet or triplet scalar multiplet, and thus 
existence of beyond the Standard Model physics.  So far LHC has searched for a charged Higgs boson decaying into $\tau\nu$ and $t \bar{b}$ which are 
good channels for a doublet like Higgs coupled to the fermions. To resolve the issue of the existence charged Higgs boson and its role
in electroweak symmetry breaking one has to look for all possible channels. 

\section*{Acknowledgement} 
PB wants to thank University of Helsinki and The Institute of Mathematical Sciences for the visits during the collaboration.
KH acknowledges the H2020-MSCA-RICE-2014 grant no. 645722 (NonMinimalHiggs).
%

%%%%%%%%%%%%%%%%%%%%%%%%%%%%%%%%%


\begin{thebibliography}{99}

%  --1 
\bibitem{ATLAS}
%\cite{Aad:2014eva}
%\bibitem{Aad:2014eva}
ATLAS-CONF-2015-007;
  G.~Aad {\it et al.}  [ATLAS Collaboration],
  %``Measurements of Higgs boson production and couplings in the four-lepton channel in pp collisions at center-of-mass energies of 7 and 8 TeV with the ATLAS detector,''
  Phys.\ Rev.\ D {\bf 91} (2015) 1,  012006
  [arXiv:1408.5191 [hep-ex]];
  %%CITATION = ARXIV:1408.5191;%%
  %52 citations counted in INSPIRE as of 24 Apr 2015
  %%\cite{ATLAS:2014aga}
%\bibitem{ATLAS:2014aga}
  G.~Aad {\it et al.}  [ATLAS Collaboration],
  %``Observation and measurement of Higgs boson decays to $WW^{\ast}$ with the ATLAS detector,''
  arXiv:1412.2641 [hep-ex];
%%CITATION = ARXIV:1412.2641;%%
  %24 citations counted in INSPIRE as of 24 Apr 2015
  %\cite{Aad:2014aba}
%\bibitem{Aad:2014aba}
  G.~Aad {\it et al.}  [ATLAS Collaboration],
  %``Measurement of the Higgs boson mass from the $H\rightarrow \gamma\gamma$ and $H \rightarrow ZZ^{*} \rightarrow 4\ell$ channels with the ATLAS detector using 25 fb$^{-1}$ of $pp$ collision data,''
  Phys.\ Rev.\ D {\bf 90} (2014) 5,  052004
  [arXiv:1406.3827 [hep-ex]].
  %%CITATION = ARXIV:1406.3827;%%
  %158 citations counted in INSPIRE as of 24 Apr 2015

% -- 2
\bibitem{CMS}
%\cite{Aad:2015zhl}
%\bibitem{Aad:2015zhl}
  G.~Aad {\it et al.}  [ATLAS and CMS Collaborations],
  %``Combined Measurement of the Higgs Boson Mass in $pp$ Collisions at $\sqrt{s}=7$ and 8 TeV with the ATLAS and CMS Experiments,''
  arXiv:1503.07589 [hep-ex];
  %%CITATION = ARXIV:1503.07589;%%
  %33 citations counted in INSPIRE as of 13 May 2015
%\cite{Khachatryan:2014jba}
%\cite{Chatrchyan:2013iaa}
%\bibitem{Chatrchyan:2013iaa}
  S.~Chatrchyan {\it et al.}  [CMS Collaboration],
  %``Measurement of Higgs boson production and properties in the WW decay channel with leptonic final states,''
  JHEP {\bf 1401} (2014) 096
  [arXiv:1312.1129 [hep-ex]];
  %%CITATION = ARXIV:1312.1129;%%
  %149 citations counted in INSPIRE as of 13 May 2015
  %\cite{Chatrchyan:2013mxa}
%\bibitem{Chatrchyan:2013mxa}
  S.~Chatrchyan {\it et al.}  [CMS Collaboration],
  %``Measurement of the properties of a Higgs boson in the four-lepton final state,''
  Phys.\ Rev.\ D {\bf 89} (2014) 9,  092007
  [arXiv:1312.5353 [hep-ex]].
  %%CITATION = ARXIV:1312.5353;%%
  %227 citations counted in INSPIRE as of 13 May 2015
  
  %--3
\bibitem{CMS2}
  V.~Khachatryan {\it et al.}  [CMS Collaboration],
  %``Precise determination of the mass of the Higgs boson and tests of compatibility of its couplings with the standard model predictions using proton collisions at 7 and 8 TeV,''
  arXiv:1412.8662 [hep-ex];
  %%CITATION = ARXIV:1412.8662;%%
  %79 citations counted in INSPIRE as of 24 Apr 2015
  %\cite{CMS:xwa}
%\bibitem{CMS:xwa}
  [CMS Collaboration],
  %``Properties of the Higgs-like boson in the decay H to ZZ to 4l in pp collisions at sqrt s =7 and 8 TeV,''
  CMS-PAS-HIG-13-002.
  %%CITATION = CMS-PAS-HIG-13-002;%%
  %228 citations counted in INSPIRE as of 13 May 2015
  
  
 % --4
\bibitem{hierarchy}
  %\cite{Witten:1981nf}
%\bibitem{Witten:1981nf}
  E.~Witten,
  %``Dynamical Breaking of Supersymmetry,''
  Nucl.\ Phys.\ B {\bf 188} (1981) 513;
  %%CITATION = NUPHA,B188,513;%%
  %2547 citations counted in INSPIRE as of 13 May 2015
  %\cite{Dimopoulos:1981zb}
%\bibitem{Dimopoulos:1981zb}
  S.~Dimopoulos and H.~Georgi,
  %``Softly Broken Supersymmetry and SU(5),''
  Nucl.\ Phys.\ B {\bf 193} (1981) 150;
  %%CITATION = NUPHA,B193,150;%%
  %2192 citations counted in INSPIRE as of 13 May 2015
  %\cite{Witten:1981kv}
%\bibitem{Witten:1981kv}
  E.~Witten,
  %``Mass Hierarchies in Supersymmetric Theories,''
  Phys.\ Lett.\ B {\bf 105} (1981) 267;
  %%CITATION = PHLTA,B105,267;%%
  %505 citations counted in INSPIRE as of 13 May 2015
  %\cite{Kaul:1981hi}
%\bibitem{Kaul:1981hi}
  R.~K.~Kaul and P.~Majumdar,
  %``Cancellation of Quadratically Divergent Mass Corrections in Globally Supersymmetric Spontaneously Broken Gauge Theories,''
  Nucl.\ Phys.\ B {\bf 199} (1982) 36;
  %%CITATION = NUPHA,B199,36;%%
  %190 citations counted in INSPIRE as of 13 May 2015
  %\cite{Sakai:1981gr}
%\bibitem{Sakai:1981gr}
  N.~Sakai,
  %``Naturalness in Supersymmetric Guts,''
  Z.\ Phys.\ C {\bf 11} (1981) 153.
  %%CITATION = ZEPYA,C11,153;%%
  %1176 citations counted in INSPIRE as of 13 May 2015

% --5
   \bibitem{LEPb}
  %\cite{Barate:2003sz}
%\bibitem{Barate:2003sz}
  R.~Barate {\it et al.}  [LEP Working Group for Higgs boson searches and ALEPH and DELPHI and L3 and OPAL Collaborations],
  %``Search for the standard model Higgs boson at LEP,''
  Phys.\ Lett.\ B {\bf 565} (2003) 61
  [hep-ex/0306033];
  %%CITATION = HEP-EX/0306033;%%
  %2206 citations counted in INSPIRE as of 15 Apr 2015
  %\cite{Schael:2006cr}
%\bibitem{Schael:2006cr}
  S.~Schael {\it et al.}  [ALEPH and DELPHI and L3 and OPAL and LEP Working Group for Higgs Boson Searches Collaborations],
  %``Search for neutral MSSM Higgs bosons at LEP,''
  Eur.\ Phys.\ J.\ C {\bf 47} (2006) 547
  [hep-ex/0602042].
  %%CITATION = HEP-EX/0602042;%%
  %697 citations counted in INSPIRE as of 15 Apr 2015
  %\cite{Aad:2015pfx}
  
  % --6
  
    \bibitem{cMSSMb}
  %\cite{Baer:2012uya}
%\bibitem{Baer:2012uya}
  H.~Baer, V.~Barger and A.~Mustafayev,
  %``Neutralino dark matter in mSUGRA/CMSSM with a 125 GeV light Higgs scalar,''
  JHEP {\bf 1205} (2012) 091
  [arXiv:1202.4038 [hep-ph]];
  %%CITATION = ARXIV:1202.4038;%%
  %109 citations counted in INSPIRE as of 13 May 2015
  %\cite{Ellis:2012aa}
%\bibitem{Ellis:2012aa}
  J.~Ellis and K.~A.~Olive,
  %``Revisiting the Higgs Mass and Dark Matter in the CMSSM,''
  Eur.\ Phys.\ J.\ C {\bf 72} (2012) 2005
  [arXiv:1202.3262 [hep-ph]];
  %%CITATION = ARXIV:1202.3262;%%
  %124 citations counted in INSPIRE as of 13 May 2015
  %\cite{Nath:2012nh}
%\bibitem{Nath:2012nh}
  P.~Nath,
  %``Higgs Physics and Supersymmetry,''
  Int.\ J.\ Mod.\ Phys.\ A {\bf 27} (2012) 1230029
  [arXiv:1210.0520 [hep-ph]].
  %%CITATION = ARXIV:1210.0520;%%
  %19 citations counted in INSPIRE as of 13 May 2015

  % --7
\bibitem{cMSSM}
   %\cite{Chamseddine:1982jx}
%\bibitem{Chamseddine:1982jx}
  A.~H.~Chamseddine, R.~L.~Arnowitt and P.~Nath,
  %``Locally Supersymmetric Grand Unification,''
  Phys.\ Rev.\ Lett.\  {\bf 49} (1982) 970;
  %%CITATION = PRLTA,49,970;%%
  %1701 citations counted in INSPIRE as of 13 May 2015 
  %\cite{Kane:1993td}
%\bibitem{Kane:1993td}
  G.~L.~Kane, C.~F.~Kolda, L.~Roszkowski and J.~D.~Wells,
  %``Study of constrained minimal supersymmetry,''
  Phys.\ Rev.\ D {\bf 49} (1994) 6173
  [hep-ph/9312272].
  %%CITATION = HEP-PH/9312272;%%
  %633 citations counted in INSPIRE as of 13 May 2015
  
  % -- 8
  %\cite{Ellwanger:2009dp}
\bibitem{Ellwanger:2009dp} 
  U.~Ellwanger, C.~Hugonie and A.~M.~Teixeira,
  %``The Next-to-Minimal Supersymmetric Standard Model,''
  Phys.\ Rept.\  {\bf 496}, 1 (2010)
  %doi:10.1016/j.physrep.2010.07.001
  [arXiv:0910.1785 [hep-ph]].
  %%CITATION = doi:10.1016/j.physrep.2010.07.001;%%
  %592 citations counted in INSPIRE as of 03 Dec 2015l
  
  % --9
  
  \bibitem{nmssm}
  
  %\cite{Miller:2003ay}
%\bibitem{Miller:2003ay} 
  D.~J.~Miller, R.~Nevzorov and P.~M.~Zerwas,
  %``The Higgs sector of the next-to-minimal supersymmetric standard model,''
  Nucl.\ Phys.\ B {\bf 681}, 3 (2004)
  %doi:10.1016/j.nuclphysb.2003.12.021
  [hep-ph/0304049];
  %%CITATION = doi:10.1016/j.nuclphysb.2003.12.021;%%
  %188 citations counted in INSPIRE as of 04 Dec 2015l
  
  %\cite{Ross:2011xv}
%\bibitem{Ross:2011xv}
  G.~G.~Ross and K.~Schmidt-Hoberg,
  %``The Fine-Tuning of the Generalised NMSSM,''
  Nucl.\ Phys.\ B {\bf 862} (2012) 710
  [arXiv:1108.1284 [hep-ph]];
  %%CITATION = ARXIV:1108.1284;%%
  %86 citations counted in INSPIRE as of 04 Jun 2015
  %\cite{Ellwanger:2011sk}
%\bibitem{Ellwanger:2011sk}
  U.~Ellwanger,
  %``Higgs Bosons in the Next-to-Minimal Supersymmetric Standard Model at the LHC,''
  Eur.\ Phys.\ J.\ C {\bf 71} (2011) 1782
  [arXiv:1108.0157 [hep-ph]];
  %%CITATION = ARXIV:1108.0157;%%
  %41 citations counted in INSPIRE as of 04 juin 2015
   %\cite{Ellwanger:2009dp}
   
    %\cite{Jeong:2011jk}
%\bibitem{Jeong:2011jk}
  K.~S.~Jeong, Y.~Shoji and M.~Yamaguchi,
  %``Peccei-Quinn invariant extension of the NMSSM,''
  JHEP {\bf 1204} (2012) 022
  [arXiv:1112.1014 [hep-ph]];
  %%CITATION = ARXIV:1112.1014;%%
  %22 citations counted in INSPIRE as of 04 juin 2015
   
    %\cite{Bomark:2014gya}
%\bibitem{Bomark:2014gya}
  N.~E.~Bomark, S.~Moretti, S.~Munir and L.~Roszkowski,
  %``A light NMSSM pseudoscalar Higgs boson at the LHC redux,''
  JHEP {\bf 1502} (2015) 044
  [arXiv:1409.8393 [hep-ph]].
  %%CITATION = ARXIV:1409.8393;%%
  %11 citations counted in INSPIRE as of 04 juin 2015

   % 10 
   
   
   %\bibitem{chiggs-xsec}
   %\cite{Bawa:1989pc}
  \bibitem{Bawa:1989pc} 
  A.~C.~Bawa, C.~S.~Kim and A.~D.~Martin,
  %``Charged Higgs Production At Hadron Colliders,''
  Z.\ Phys.\ C {\bf 47}, 75 (1990);
  %doi:10.1007/BF01551915
  %%CITATION = doi:10.1007/BF01551915;%%
  %96 citations counted in INSPIRE as of 04 Dec 2015
  
  %\cite{Datta:2001qs}
  \bibitem{Datta:2001qs} 
  A.~Datta, A.~Djouadi, M.~Guchait and Y.~Mambrini,
  %``Charged Higgs production from SUSY particle cascade decays at the CERN LHC,''
  Phys.\ Rev.\ D {\bf 65}, 015007 (2002)
  %doi:10.1103/PhysRevD.65.015007
  [hep-ph/0107271].
  %%CITATION = doi:10.1103/PhysRevD.65.015007;%%
  %45 citations counted in INSPIRE as of 04 Dec 2015

  % 11
  %\cite{Gunion:1993sv}
\bibitem{Gunion:1993sv} 
  J.~F.~Gunion,
  %``Detecting the t b decays of a charged Higgs boson at a hadron supercollider,''
  Phys.\ Lett.\ B {\bf 322}, 125 (1994)
  %doi:10.1016/0370-2693(94)90500-2
  [hep-ph/9312201].
  %%CITATION = doi:10.1016/0370-2693(94)90500-2;%%
  %92 citations counted in INSPIRE as of 04 Dec 2015
  
   \bibitem{NMSSMps}

    %\bibitem{Andreas:2010ms}
  S.~Andreas, O.~Lebedev, S.~Ramos-Sanchez and A.~Ringwald,
  %``Constraints on a very light CP-odd Higgs of the NMSSM and other axion-like particles,''
  JHEP {\bf 1008} (2010) 003
  [arXiv:1005.3978 [hep-ph]];
  %%CITATION = ARXIV:1005.3978;%%
  %31 citations counted in INSPIRE as of 04 Jun 2015
  
%\cite{Cacciapaglia:2013ora}
%\bibitem{Cacciapaglia:2013ora}
  G.~Cacciapaglia, A.~Deandrea, G.~D.~La Rochelle and J.~B.~Flament,
  %``Searching for a lighter Higgs boson: Parametrization and sample tests,''
  Phys.\ Rev.\ D {\bf 91} (2015) 1,  015012
  [arXiv:1311.5132 [hep-ph]];
  %%CITATION = ARXIV:1311.5132;%%
  %6 citations counted in INSPIRE as of 13 Oct 2015
  %\bibitem{Christensen:2013dra}
  N.~D.~Christensen, T.~Han, Z.~Liu and S.~Su,
  %``Low-Mass Higgs Bosons in the NMSSM and Their LHC Implications,''
  JHEP {\bf 1308} (2013) 019
  [arXiv:1303.2113 [hep-ph]];
  %%CITATION = ARXIV:1303.2113;%%
  %37 citations counted in INSPIRE as of 13 Oct 2015
  %\cite{King:2014xwa}
 %\bibitem{King:2014xwa} 
  S.~F.~King, M.~M�hlleitner, R.~Nevzorov and K.~Walz,
  %``Discovery Prospects for NMSSM Higgs Bosons at the High-Energy Large Hadron Collider,''
  Phys.\ Rev.\ D {\bf 90}, no. 9, 095014 (2014)
  %doi:10.1103/PhysRevD.90.095014
  [arXiv:1408.1120 [hep-ph]];
  %%CITATION = doi:10.1103/PhysRevD.90.095014;%%
  %30 citations counted in INSPIRE as of 08 Dec 2015
% %\cite{Cao:2015loa}
% %\bibitem{Cao:2015loa}
%   J.~Cao, L.~Shang, P.~Wu, J.~M.~Yang and Y.~Zhang,
%   %``Interpreting the galactic center gamma-ray excess in the NMSSM,''
%   JHEP {\bf 1510} (2015) 030
%   [arXiv:1506.06471 [hep-ph]].
%   %%CITATION = ARXIV:1506.06471;%%
%   %9 citations counted in INSPIRE as of 13 Oct 2015

%\cite{Guchait:2015owa}
%\bibitem{Guchait:2015owa}
  M.~Guchait and J.~Kumar,
  %``Light Higgs Bosons in NMSSM at the LHC,''
  arXiv:1509.02452 [hep-ph].
  %%CITATION = ARXIV:1509.02452;%%
  %1 citations counted in INSPIRE as of 13 Oct 2015


  % 12
  %\cite{Rathsman:2012dp}
\bibitem{Rathsman:2012dp} 
  J.~Rathsman and T.~Rossler,
  %``Closing the Window on Light Charged Higgs Bosons in the NMSSM,''
  Adv.\ High Energy Phys.\  {\bf 2012}, 853706 (2012)
  %doi:10.1155/2012/853706
  [arXiv:1206.1470 [hep-ph]].
  %%CITATION = doi:10.1155/2012/853706;%%
  %16 citations counted in INSPIRE as of 03 Dec 2015
  
   % 13
  %\cite{Coleppa:2014cca}
  \bibitem{chargedhiggs-pheno}
%\bibitem{Coleppa:2014cca} 
  B.~Coleppa, F.~Kling and S.~Su,
  %``Charged Higgs search via $AW^\pm/HW^\pm$ channel,''
  JHEP {\bf 1412}, 148 (2014)
  %doi:10.1007/JHEP12(2014)148
  [arXiv:1408.4119 [hep-ph]];
  %%CITATION = doi:10.1007/JHEP12(2014)148;%%
  %11 citations counted in INSPIRE as of 03 Dec 2015
  
  %\cite{Kling:2015uba}
%\bibitem{Kling:2015uba} 
  F.~Kling, A.~Pyarelal and S.~Su,
  %``Light Charged Higgs Bosons to AW/HW via Top Decay,''
  JHEP {\bf 1511}, 051 (2015)
  %doi:10.1007/JHEP11(2015)051
  [arXiv:1504.06624 [hep-ph]].
  %%CITATION = doi:10.1007/JHEP11(2015)051;%%
  %3 citations counted in INSPIRE as of 17 f�vr. 2016

  \bibitem{nmssmtools}
 %\cite{Ellwanger:2004xm}
%\bibitem{Ellwanger:2004xm} 
  U.~Ellwanger, J.~F.~Gunion and C.~Hugonie,
  %``NMHDECAY: A Fortran code for the Higgs masses, couplings and decay widths in the NMSSM,''
  JHEP {\bf 0502}, 066 (2005)
  %doi:10.1088/1126-6708/2005/02/066
  [hep-ph/0406215];
  %%CITATION = doi:10.1088/1126-6708/2005/02/066;%%
  %312 citations counted in INSPIRE as of 17 f�vr. 2016
  
  %\cite{Ellwanger:2005dv}
%\bibitem{Ellwanger:2005dv} 
  U.~Ellwanger and C.~Hugonie,
  %``NMHDECAY 2.0: An Updated program for sparticle masses, Higgs masses, couplings and decay widths in the NMSSM,''
  Comput.\ Phys.\ Commun.\  {\bf 175}, 290 (2006)
  %doi:10.1016/j.cpc.2006.04.004
  [hep-ph/0508022];
  %%CITATION = doi:10.1016/j.cpc.2006.04.004;%%
  %287 citations counted in INSPIRE as of 17 f�vr. 2016

  %\cite{Belanger:2005kh}
%\bibitem{Belanger:2005kh} 
  G.~Belanger, F.~Boudjema, C.~Hugonie, A.~Pukhov and A.~Semenov,
  %``Relic density of dark matter in the NMSSM,''
  JCAP {\bf 0509}, 001 (2005)
  %doi:10.1088/1475-7516/2005/09/001
  [hep-ph/0505142].
  %%CITATION = doi:10.1088/1475-7516/2005/09/001;%%
  %152 citations counted in INSPIRE as of 17 Feb 2016

  % 15
    %\cite{Heinemeyer:2013tqa}
  \bibitem{chargino}
 See: \url{http://lepsusy.web.cern.ch/lepsusy/www/inos_moriond01/charginos_pub.html}
 
 
 % 16
 
  \bibitem{hiddenHE}
  %\cite{Bandyopadhyay:2015tva}
%\bibitem{Bandyopadhyay:2015tva}
  P.~Bandyopadhyay, C.~Coriano and A.~Costantini,
  %``Probing the Hidden Higgs Bosons of the $Y=0$ Triplet- and Singlet-Extended Supersymmetric Standard Model at the LHC,''
  arXiv:1510.06309 [hep-ph];
  %%CITATION = ARXIV:1510.06309;%%
  %\cite{Bandyopadhyay:2015oga}
%\bibitem{Bandyopadhyay:2015oga}
  P.~Bandyopadhyay, C.~Coriano and A.~Costantini,
  %``Perspectives on a supersymmetric extension of the standard model with a Y = 0 Higgs triplet and a singlet at the LHC,''
  JHEP {\bf 1509} (2015) 045
  [arXiv:1506.03634 [hep-ph]].
  %%CITATION = ARXIV:1506.03634;%%
  %1 citations counted in INSPIRE as of 30 Oct 2015
 \bibitem{thridgensusy}
  G.~Aad {\it et al.} [ATLAS Collaboration],
  %``ATLAS Run 1 searches for direct pair production of third-generation squarks at the Large Hadron Collider,''
  arXiv:1506.08616 [hep-ex];
  %%CITATION = ARXIV:1506.08616;%%
  %21 citations counted in INSPIRE as of 13 Oct 2015
  %\cite{Khachatryan:2015wza}
%\bibitem{Khachatryan:2015wza}
  V.~Khachatryan {\it et al.} [CMS Collaboration],
  %``Searches for third-generation squark production in fully hadronic final states in proton-proton collisions at $ \sqrt{s} = 8$ TeV,''
  JHEP {\bf 1506} (2015) 116
  [arXiv:1503.08037 [hep-ex]].
  %%CITATION = ARXIV:1503.08037;%%
  %23 citations counted in INSPIRE as of 13 Oct 2015
 
% \bibitem{rparity}
%     %\cite{Barbier:2004ez}
% %\bibitem{Barbier:2004ez}
%   R.~Barbier, C.~Berat, M.~Besancon, M.~Chemtob, A.~Deandrea, E.~Dudas, P.~Fayet and S.~Lavignac, G.~Moreau, E.~Perez, Y.~Sirois
%   %``R-parity violating supersymmetry,''
%   Phys.\ Rept.\  {\bf 420} (2005) 1
%   [hep-ph/0406039].
%   %%CITATION = HEP-PH/0406039;%%
%   %688 citations counted in INSPIRE as of 13 May 2015

%  \bibitem{pMSSMb}
%  %\cite{Carena:2011aa}
% %\bibitem{Carena:2011aa}
%   M.~Carena, S.~Gori, N.~R.~Shah and C.~E.~M.~Wagner,
%   %``A 125 GeV SM-like Higgs in the MSSM and the $\gamma \gamma$ rate,''
%   JHEP {\bf 1203} (2012) 014
%   [arXiv:1112.3336 [hep-ph]].
%   %%CITATION = ARXIV:1112.3336;%%
%   %314 citations counted in INSPIRE as of 13 May 2015
%  %\cite{Hall:2011aa}
% %\bibitem{Hall:2011aa}
%   L.~J.~Hall, D.~Pinner and J.~T.~Ruderman,
%   %``A Natural SUSY Higgs Near 126 GeV,''
%   JHEP {\bf 1204} (2012) 131
%   [arXiv:1112.2703 [hep-ph]].
%   %%CITATION = ARXIV:1112.2703;%%
%   %364 citations counted in INSPIRE as of 13 May 2015
%   %\cite{Arbey:2011ab}
% %\bibitem{Arbey:2011ab}
%   A.~Arbey, M.~Battaglia, A.~Djouadi, F.~Mahmoudi and J.~Quevillon,
%   %``Implications of a 125 GeV Higgs for supersymmetric models,''
%   Phys.\ Lett.\ B {\bf 708} (2012) 162
%   [arXiv:1112.3028 [hep-ph]].
%   %%CITATION = ARXIV:1112.3028;%%
%   %259 citations counted in INSPIRE as of 13 May 2015

 \bibitem{CTEQ}
    J.~Pumplin, D.~R.~Stump, J.~Huston, H.~L.~Lai, P.~Nadolsky and W.~K.~Tung,
  JHEP {\bf 0207}, 012 (2002)
  [arXiv:hep-ph/0201195].
  %%CITATION = JHEPA,0207,012;%%
  
% 17
\bibitem{sarah}
  F.~Staub,
  %``SARAH 3.2: Dirac Gauginos, UFO output, and more,''
  Comput.\ Phys.\ Commun.\  {\bf 184} (2013) pp. 1792
   [Comput.\ Phys.\ Commun.\  {\bf 184} (2013) 1792]
  [arXiv:1207.0906 [hep-ph]].
  %%CITATION = ARXIV:1207.0906;%%
  %70 citations counted in INSPIRE as of 24 Apr 2015

  % 18
    \bibitem{calchep}
  A.~Pukhov,
``CalcHEP 3.2: MSSM, structure functions, event generation, batchs, and
generation of matrix elements for other packages'',
  [arXiv:hep-ph/0412191].
%CITATION = HEP-PH/0412191;
   
 % 19
%\cite{Sjostrand:2001yu}
\bibitem{pythia}
  T.~Sjostrand, L.~Lonnblad and S.~Mrenna,
% ``PYTHIA 6.2: Physics and manual,''
  [arXiv:hep-ph/0108264].
% CITATION = HEP-PH/0108264;
%\cite{Lee:2007gn}
%%%
 % 20
%\cite{Lai:1999wy}
\bibitem{slha}
  P.~Skands {\it et al.},
  %``SUSY Les Houches accord: Interfacing SUSY spectrum calculators, decay
  %packages, and event generators,''
  JHEP {\bf 0407}, 036 (2004)
  [arXiv:hep-ph/0311123]; \\ see also
\url{http://skands.physics.monash.edu/slha/}.
  %%CITATION = JHEPA,0407,036;%%
%\cite{Sjostrand:2001yu}

  \bibitem{Kidonakis:2008vd} 
  N.~Kidonakis,
  %``Higher order corrections to H+- production,''
  PoS CHARGED {\bf 2008}, 003 (2008)
  [arXiv:0811.4757 [hep-ph]].
  %%CITATION = ARXIV:0811.4757;%%
  %6 citations counted in INSPIRE as of 04 Dec 2015


\bibitem{fastjet}
  M.~Cacciari, G.~P.~Salam and G.~Soyez,
  %``FastJet User Manual,''
  Eur.\ Phys.\ J.\ C {\bf 72} (2012) 1896
  [arXiv:1111.6097 [hep-ph]].
  %%CITATION = ARXIV:1111.6097;%%
  %1019 citations counted in INSPIRE as of 13 Oct 2015
  
  \bibitem{ca-algo} 
  Y.~L.~Dokshitzer, G.~D.~Leder, S.~Moretti and B.~R.~Webber,
  %``Better jet clustering algorithms,''
  JHEP {\bf 9708}, 001 (1997)
  %doi:10.1088/1126-6708/1997/08/001
  [hep-ph/9707323];
  %%CITATION = doi:10.1088/1126-6708/1997/08/001;%%
  %603 citations counted in INSPIRE as of 17 Dec 2015

    \bibitem{taujet}
%\cite{Bagliesi:2007qx}
%\bibitem{Bagliesi:2007qx}
  G.~Bagliesi,
  %``Tau tagging at Atlas and CMS,''
  arXiv:0707.0928 [hep-ex];
  %%CITATION = ARXIV:0707.0928;%%
  %7 citations counted in INSPIRE as of 11 Nov 2014
  
    \bibitem{btag}
%\cite{Tomalin:2007zz}
%\bibitem{Tomalin:2007zz}
  I.~R.~Tomalin [CMS Collaboration],
  %``b tagging in CMS,''
  J.\ Phys.\ Conf.\ Ser.\  {\bf 110} (2008) 092033.
  %%CITATION = 00462,110,092033;%%
  %2 citations counted in INSPIRE as of 13 Oct 2015

\bibitem{alpgen}
%\cite{Mangano:2002ea}
%\bibitem{Mangano:2002ea}
M.~L.~Mangano, M.~Moretti, F.~Piccinini, R.~Pittau and A.~D.~Polosa,
%``ALPGEN, a generator for hard multiparton processes in hadronic collisions,''
JHEP {\bf 0307}, 001 (2003)
doi:10.1088/1126-6708/2003/07/001
[hep-ph/0206293].
%%CITATION = doi:10.1088/1126-6708/2003/07/001;%%
%2713 citations counted in INSPIRE as of 23 Apr 2016
  %\cite{Dokshitzer:1997in}

%\cite{Hoche:2006ph}
\bibitem{Hoche:2006ph} 
  S.~Hoeche, F.~Krauss, N.~Lavesson, L.~Lonnblad, M.~Mangano, A.~Schalicke and S.~Schumann,
  %``Matching parton showers and matrix elements,''
  hep-ph/0602031.
  %%CITATION = HEP-PH/0602031;%%
  %274 citations counted in INSPIRE as of 23 Apr 2016
  \bibitem{taumistag}
CMS PAS TAU-11-001, See also,
\url{https://inspirehep.net/record/925248/files/TAU-11-001-pas.pdf}

%\cite{Wobisch:1998wt}
%\bibitem{Wobisch:1998wt} 
  M.~Wobisch and T.~Wengler,
  %``Hadronization corrections to jet cross-sections in deep inelastic scattering,''
  In *Hamburg 1998/1999, Monte Carlo generators for HERA physics* 270-279
  [hep-ph/9907280].
  %%CITATION = HEP-PH/9907280;%%
  %337 citations counted in INSPIRE as of 17 Dec 2015
  
 \bibitem{ChATLAS}
  %\cite{Aad:2014kga}
%\bibitem{Aad:2014kga}
  G.~Aad {\it et al.} [ATLAS Collaboration],
  %``Search for charged Higgs bosons decaying via $H^{\pm} \rightarrow \tau^{\pm}\nu$ in fully hadronic final states using $pp$ collision data at $\sqrt{s} = 8$ TeV with the ATLAS detector,''
  JHEP {\bf 1503} (2015) 088
  [arXiv:1412.6663 [hep-ex]];
  
  %\cite{Aad:2015typ}
 %\bibitem{Aad:2015typ} 
  G.~Aad {\it et al.} [ATLAS Collaboration],
  %``Search for charged Higgs bosons in the $H^{\pm} \rightarrow tb$ decay channel in $pp$ collisions at $\sqrt{s} = 8$ TeV using the ATLAS detector,''
  arXiv:1512.03704 [hep-ex].
  %%CITATION = ARXIV:1512.03704;%%

  \bibitem{ChCMS}
  %\cite{CMS:2014cdp}
%\bibitem{CMS:2014cdp}
  CMS Collaboration [CMS Collaboration],
  %``Search for charged Higgs bosons with the H+ to tau nu decay channel in the fully hadronic final state at sqrt s = 8 TeV,''
  CMS-PAS-HIG-14-020;
  %%CITATION = CMS-PAS-HIG-14-020;%%
  %41 citations counted in INSPIRE as of 13 Oct 2015
  %\cite{CMS:2014pea}
%\bibitem{CMS:2014pea}
  CMS Collaboration [CMS Collaboration],
  %``Search for a heavy charged Higgs boson in proton-proton collisions at sqrts=8 TeV with the CMS detector,''
  CMS-PAS-HIG-13-026.
  %%CITATION = CMS-PAS-HIG-13-026;%%
  %27 citations counted in INSPIRE as of 13 Oct 2015

 
    %\cite{Ball:2007zza}
%\bibitem{Ball:2007zza}
  G.~L.~Bayatian {\it et al.}  [CMS Collaboration],
  %``CMS technical design report, volume II: Physics performance,''
  J.\ Phys.\ G {\bf 34} (2007) 995.
  %%CITATION = JPAGA,G34,995;%%
  %1136 citations counted in INSPIRE as of 11 Nov 2014 
  
  \bibitem{CPVMSSM}
    %\cite{Bandyopadhyay:2010tv}
    %\cite{Bandyopadhyay:2011vc}
%\bibitem{Bandyopadhyay:2011vc}
  P.~Bandyopadhyay and K.~Huitu,
  %``Production of two higgses at the Large Hadron Collider in CP-violating MSSM,''
  JHEP {\bf 1311} (2013) 058
  [arXiv:1106.5108 [hep-ph]];
  %%CITATION = ARXIV:1106.5108;%%
  %4 citations counted in INSPIRE as of 14 Oct 2015
%\bibitem{Bandyopadhyay:2010tv}
  P.~Bandyopadhyay,
  %``Higgs production in CP-violating supersymmetric cascade decays: Probing the `open hole' at the Large Hadron Collider,''
  JHEP {\bf 1108} (2011) 016
  [arXiv:1008.3339 [hep-ph]];
  %%CITATION = ARXIV:1008.3339;%%
  %13 citations counted in INSPIRE as of 14 Oct 2015
  %\cite{Bandyopadhyay:2007cp}
%\bibitem{Bandyopadhyay:2007cp}
  P.~Bandyopadhyay, A.~Datta, A.~Datta and B.~Mukhopadhyaya,
  %``Associated Higgs production in CP-violating supersymmetry: Probing the 'open hole' at the large hadron collider,''
  Phys.\ Rev.\ D {\bf 78} (2008) 015017
  [arXiv:0710.3016 [hep-ph]];
  %%CITATION = ARXIV:0710.3016;%%
  %17 citations counted in INSPIRE as of 14 Oct 2015
  %\cite{Pilaftsis:1999qt}
  %\cite{Ghosh:2004wr}
%\bibitem{Ghosh:2004wr}
  D.~K.~Ghosh and S.~Moretti,
  %``Probing the light neutral Higgs boson scenario of the CP-violating MSSM Higgs sector at the LHC,''
  Eur.\ Phys.\ J.\ C {\bf 42} (2005) 341
  [hep-ph/0412365];
  %%CITATION = HEP-PH/0412365;%%
  %19 citations counted in INSPIRE as of 14 Oct 2015
  %\cite{Ghosh:2004cc}
 %\cite{Ghosh:2004cc}
%\bibitem{Ghosh:2004cc}
  D.~K.~Ghosh, R.~M.~Godbole and D.~P.~Roy,
  %``Probing the CP-violating light neutral Higgs in the charged Higgs decay at the LHC,''
  Phys.\ Lett.\ B {\bf 628} (2005) 131
  [hep-ph/0412193].
  %%CITATION = HEP-PH/0412193;%%
  %38 citations counted in INSPIRE as of 14 Oct 2015

  \bibitem{tripch}

  %\cite{Bandyopadhyay:2014vma}
%\bibitem{Bandyopadhyay:2014vma}
  P.~Bandyopadhyay, K.~Huitu and A.~S.~Keceli,
  %``Multi-Lepton Signatures of the Triplet Like Charged Higgs at the LHC,''
  JHEP {\bf 1505} (2015) 026
  [arXiv:1412.7359 [hep-ph]];
  %%CITATION = ARXIV:1412.7359;%%
  %\cite{Bandyopadhyay:2014tha}
%\bibitem{Bandyopadhyay:2014tha}
  P.~Bandyopadhyay, S.~Di Chiara, K.~Huitu and A.~S.~Keceli,
  %``Naturality vs perturbativity, B$_{s}$ physics, and LHC data in triplet extension of MSSM,''
  JHEP {\bf 1411} (2014) 062
  [arXiv:1407.4836 [hep-ph]];
  %%CITATION = ARXIV:1407.4836;%%
  %9 citations counted in INSPIRE as of 13 Oct 2015
  %\cite{Bandyopadhyay:2013lca}
%\bibitem{Bandyopadhyay:2013lca}
  P.~Bandyopadhyay, K.~Huitu and A.~Sabanci,
  %``Status of $Y=0$ Triplet Higgs with supersymmetry in the light of $\sim 125$ GeV Higgs discovery,''
  JHEP {\bf 1310} (2013) 091
  [arXiv:1306.4530 [hep-ph]].
  %%CITATION = ARXIV:1306.4530;%%
  %11 citations counted in INSPIRE as of 13 Oct 2015
%\bibitem{pbancc}
%P.~Bandyopadhyay, C.~Corian\`o, A.~Costantini,
%Phenomenology of light charged Higgs boson in TNMSSM at the LHC, \emph{in preparation}.

%\cite{Bandyopadhyay:2015ifm}
\bibitem{Bandyopadhyay:2015ifm}
  P.~Bandyopadhyay, C.~Coriano and A.~Costantini,
  %``A General Analysis of the Higgs Sector of the $Y=0$ Triplet-Singlet Extension of the MSSM at the LHC,''
  arXiv:1512.08651 [hep-ph].
  %%CITATION = ARXIV:1512.08651;%%
%\cite{Agashe:2011ia}
% \bibitem{agashe}
%   K.~Agashe, A.~Azatov, A.~Katz and D.~Kim,
%   %``Improving the tunings of the MSSM by adding triplets and singlet,''
%   Phys.\ Rev.\ D {\bf 84} (2011) 115024
%   [arXiv:1109.2842 [hep-ph]].
%   %%CITATION = ARXIV:1109.2842;%%
%   %19 citations counted in INSPIRE as of 19 Jan 2015

\end{thebibliography}
\end{document}